\documentclass[prb,twocolumn,floatfix]{revtex4}

\usepackage{amsfonts}
\usepackage{amsmath}
\usepackage{verbatim}
\usepackage[dvips]{graphicx}
\usepackage{subfigure}

\def\Tr{\textrm}
\def\dd{\textrm{d}}
\def\Bf{\boldsymbol}
\def\Wt{\widetilde}

\def\rv{\Bf{r}}

\def\kv{\Bf{k}}

\def\qv{\Bf{q}}
\def\Qv{\Bf{Q}}

\begin{document}

\title{Unitarity in periodic potentials: a renormalization group analysis}
\author{Predrag Nikoli\'c$^{1,2}$}
\affiliation{$^1$Department of Physics and Astronomy, George Mason University, Fairfax, VA 22030, USA}
\affiliation{$^2$Institute for Quantum Matter at Johns Hopkins University, Baltimore, MD 21218, USA}
\date{\today}

\begin{abstract}

We explore the universal properties of interacting fermionic lattice systems, mostly focusing on the development of pairing correlations from attractive interactions. Using renormalization group we identify a large number of fixed points and show that they correspond to resonant scattering in multiple channels. Pairing resonances in finite-density band insulators occur between quasiparticles and quasiholes living at different symmetry-related wavevectors in the Brillouin zone. This allows a BCS-BEC crossover interpretation of both Cooper and particle-hole pairing. We show that in two dimensions the run-away flows of relevant attractive interactions lead to charged-boson-dominated low energy dynamics in the insulating states, and superfluid transitions in bosonic mean-field or XY universality classes. Analogous phenomena in higher dimensions are restricted to the strong coupling limit, while at weak couplings the transition is in the pair-breaking BCS class. The models discussed here can be realized with ultra-cold gases of alkali atoms tuned to a broad Feshbach resonance in an optical lattice, enabling experimental studies of pairing correlations in insulators, especially in their universal regimes. In turn, these simple and tractable models capture the emergence of fluctuation-driven superconducting transitions in fermionic systems, which is of interest in the context of high temperature superconductors.

\end{abstract}

\maketitle

\section{Introduction}

Fermionic ultra-cold atoms with nearly resonant scattering in the unitarity regime \cite{Zwierlein2005, Partridge2005, Kohl2005, Stoferle2006, Zwierlein2006, Chin2006, Partridge2006, Stewart2008, Jordens2008, Schneider2008, Gaebler2010} realize the strongest possible form of Cooper pairing, revealed by critical velocity \cite{Miller2007, Diener2008}. It is natural to expect that zero-temperature normal states near unitarity would be very strongly correlated with a host of unconventional properties, possibly bearing some resemblance to those found in cuprates \cite{Stajic2004}. The unitarity limit is therefore an excellent starting point for studies of correlated fermionic superconductors and insulators, which has not been exploited enough in literature. The benefits are both theoretical and experimental. Systematic perturbative and renormalization group calculations are feasible mainly because the unperturbed ground state (fixed point) is a simple state, the vacuum or a band insulator. Experimentally, the unitarity limit is routinely accessed in cold gases of alkali atoms tuned near a broad Feshbach resonance \cite{Bloch08p885, Tiesinga2009}.

Our ultimate goal is to address the long-standing questions about the nature of unconventional normal states proximate to strongly paired fermionic superfluids or superconductors. We design here a simple and tractable model in which the fermionic excitation gap is opened by an external periodic potential, rather than strong interactions. Attractive interactions between quasiparticles whose energy scale exceeds this gap can still give rise to pairing and superfluidity. The need for strong interactions justifies asking if the normal state proximate to the superfluid might have some unconventional properties reflecting strong correlations, especially in universal regimes shaped by resonant scattering.

The inquiry into unconventional superfluidity from the resonant scattering point of view began a long time ago \cite{Eagles1969}. The present interest in this subject is driven in parallel by a variety of unconventional superconductors in condensed matter physics, and ultra-cold atoms with nearly resonant scattering. The recent theoretical studies of scattering resonances in lattice potentials \cite{Fedichev2004, Carr2005, Dickerscheid2005, Gubbels2006, Zhou2006, Titvinidze2009, Watanabe2009} often rely on two-channel tight-binding models, featuring fermionic atoms resonantly coupled to closed channel bosonic particles. It has been argued that models of this kind provide a good effective description of the microscopic lattice systems of interest \cite{Duan2005, Diener2006, Koetsier2006}. The findings of these studies include lattice Feshbach resonances shifted from their empty-space values.

Most of the mentioned theoretical works approach resonant scattering from a somewhat microscopic angle, exemplified by perturbation theory with the vacuum unperturbed ground-state. Indeed, the usual universal behavior of particles tuned to a broad Feshbach resonance is established in the low density limit. However, universality is a many-body phenomenon and its complete description requires field-theoretical tools such as renormalization group. This issue becomes pressing in the present problem of interest, a band insulator that contains a macroscopic number of fermions in fully populated bands. These fermions may not be dynamically inert despite the Pauli exclusion principle, due to the strong interactions which bring the system to its unitarity regime.

In this paper we take a field-theoretical approach to nearly resonant pairing between gapped fermions. Abandoning all microscopic details in a renormalization group (RG) calculation allows us to gain a perspective on the generic and universal behavior of a large class of fermionic lattice systems. We analyze effective theories which are constructed to preserve the universality class of the microscopic system. This ensures that the universal phase diagram and other macroscopic properties of the microscopic system are correctly captured despite the neglect of microscopic details. The price to pay is the necessity to deal with multiple flavors of low energy quasiparticles, such as particle and hole excitations which may exist at multiple wavevectors in the first Brillouin zone. Note that the two-channel models with dynamical bosonic fields used in many previous studies may in some cases describe different physics than this paper (our interest are broad Feshbach resonances). We characterize the universality stemming from resonant scattering of quasiparticles in band insulators, and discover generalized unitarity regimes in which quasiparticles of different flavors scatter resonantly. The manifestations of unitarity which we discuss include universal ratios of measurable quantities such as critical temperature, pressure and density. We also analyze the types and conditions for pairing instabilities, conventional versus unconventional superfluid transitions, and emphasize the existence of correlated bosonic Mott insulating states in the phase diagram.

The RG analysis reveals why pairing fluctuations indeed play the crucial role in systems of gapped fermions with short-range attractive interactions. Unlike previous studies of unitarity in continuum lattice potentials, which focused on the zero-density limit \cite{Burovski2006, Zhai2007, moon:230403, Burkov2009}, we point out that the unitarity regime in the same universality class can be found at finite densities, near properly tuned transitions between the superfluid and any band insulator. The structure of fixed points depends on whether both particles and holes participate equally in the dynamics, or just one of the two quasiparticle types. In the latter case, the exact RG equations can be derived, which allows one to track the run-away flows of attractive interaction couplings (as in some studies of Iron-pnictides \cite{Wang2009, Thomale2009}). It is these run-away flows that can lead to boson-dominated dynamics at low energies. We find that instabilities in the particle-hole channel are discouraged by attractive interactions.

\begin{figure}
\includegraphics[width=2.7in]{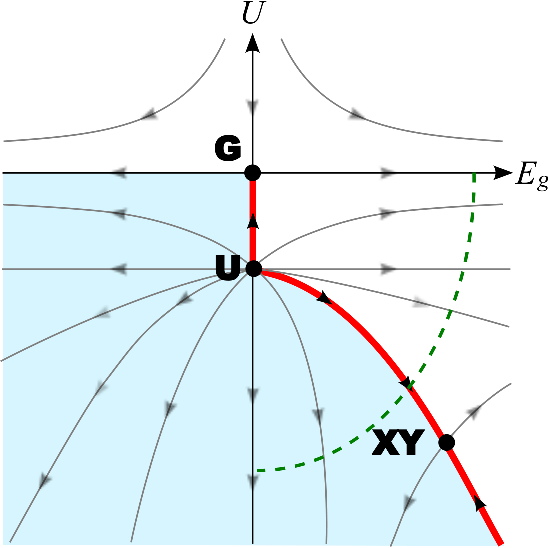}
\caption{\label{FullRG}(color online) A hypothetical renormalization group (RG) flow diagram at zero temperature of a fermionic lattice theory with attractive interactions in $d \ge 2$ dimensions. The parameters are fermion density-density interaction $U$ and bandgap (or negative chemical potential) $E_g$. This paper explores in detail the vicinity of two weak-coupling fixed points which govern the pair-breaking superfluid to insulator transition: Gaussian (G) and unitarity (U), separated in proportion to $\epsilon=d-2$. The negative couplings $U$ below unitarity experience run-away flows under RG and lead to the boson-dominated dynamics. A superfluid-insulator transition in this regime is captured by a bosonic effective theory with dynamical exponent $z=2$ (mean-field universality) or $z=1$ (XY universality). In the latter case, one expects an additional fixed point (XY) in $d \le 3$. The Gaussian fixed point of the bosonic effective theory appears at the transition line in the limit $U\to-\infty$, $E_g\to\infty$. The shaded area is the superfluid or superconducting phase, and the red thick line is the second order superfluid-insulator transition. The dashed green line encloses the region in which the fermionic RG is valid.}
\end{figure}

The run-away flows imply the ultimate RG breakdown when the diverging couplings reach cut-off scales. However, the resulting low energy bosonic dynamics is known to introduce additional fixed points associated with superfluid transitions \cite{Fisher1989a}, which appear as strong-coupling fixed points in the present RG. The superfluid transition in this regime can be either in the bosonic mean-field, or XY universality class. The mean-field universality with dynamical exponent $z=2$ emerges as the result of run-away flows from the unitarity dominated by either particles or holes, while the $z=1$ XY universality is related in the same fashion to the unitarity shaped by \emph{both} particles and holes. Therefore, the analysis here provides a glimpse of the more complete structure of fixed points in theories of fermionic particles with attractive interactions, sketched in Fig.\ref{FullRG}.

The mentioned finite-density fixed points describe unitarity in zero-density effective theories of particle and hole excitations. Therefore, one can relate a nearly critical interaction strength to scattering lengths in collisions among particles and holes. Any attractive interaction in two dimensions effectively puts the low energy quasiparticles into their Bose-Einstein condensate (BEC) limit, so quasiparticles injected in the insulating state immediately combine into bound-state pairs \cite{L1977} (whose size can be very large at weak couplings). The effective Bardeen-Cooper-Schrieffer (BCS) regime exists only above two dimensions, at least in the weak coupling limit (``above'' unitarity $|U|<|U^*|\propto\epsilon$ in Fig.\ref{FullRG}). Note that localization tendencies due to the lattice potential enhance the strength of effective interactions with respect to those between completely free fermions \cite{Fedichev2004, Koetsier2006}.

We begin the discussion by laying out the effective theory of a lattice fermionic system in the section \ref{secModel}. Then, in the section \ref{secP} we analyze the exact RG for a single species of fermions, either particles or holes, and reveal the development of short range pairing correlations in insulating states. The section \ref{secPH} presents the RG calculations for particles and holes and identifies a large number of fixed points associated with resonant scattering. This analysis is expanded in the section \ref{secMF} to multiple quasiparticle species living at different wavevectors in the Brillouin zone. All results and conclusions are summarized in the discussion section \ref{secDiscussion}.

\subsection{Model}\label{secModel}

\begin{figure}
\includegraphics[width=2.8in]{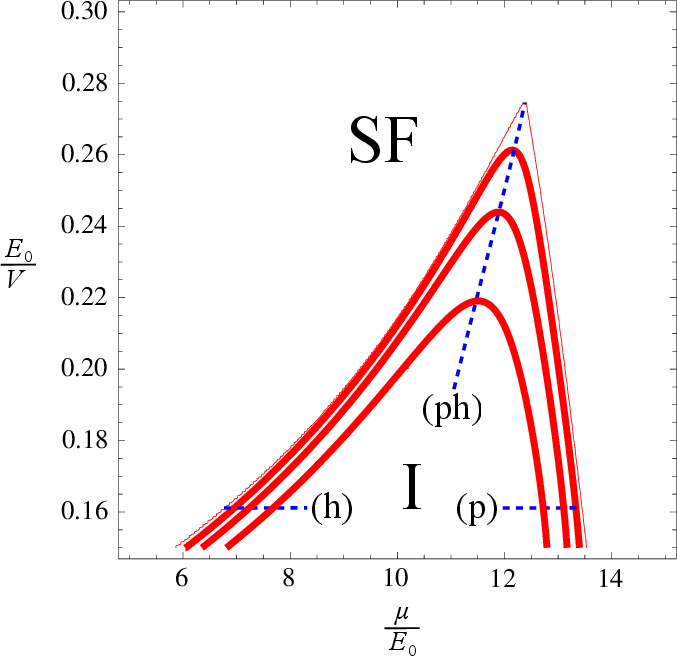}
\caption{\label{pd1}(color online) Superfluid (SF) transitions out of a two-dimensional band insulator (I) at zero temperature. The three shown transitions with thick red lines correspond to arbitrarily chosen different strengths of contact attractive interactions, becoming stronger going from top to bottom. The lattice potential is given by $V(\rv)=2V \lbrack \cos( 2\pi x / a_L ) + \cos( 2\pi y / a_L ) \rbrack$ in two dimensions, where $V$ is the lattice amplitude and $a_L$ the lattice spacing. Both $V$ and $\mu$ are measured in the units of ``recoil energy'' $E_{\Tr r}=\hbar^2/2ma_L^2$. The thin red line outlines the band edge of the corresponding non-interacting model. The dashed blue lines are trajectories in the parameter space along which the transitions dominated by particles (p), holes (h), or both (ph) can occur.}
\end{figure}

We start with a typical microscopic model which contains pairing instabilities of band insulators. Consider the second-quantized Hamiltonian of fermionic particles with short-range density-density interactions $U$, in a lattice potential $V(\rv)$:
\begin{eqnarray}\label{ContModel1}
&& \!\!\!\!\!\!\! H = \int \dd^{d}r \; \psi_{\alpha}^{\dagger}
  \left( - \frac{\nabla^2}{2m} + V(\boldsymbol{r}) - \mu \right) \psi_{\alpha} \\
&& \!\!\!\!\!\!\! +\int \dd^{d}r_1 \dd^{d}r_2 U(|\rv_1-\rv_2|)
       \psi_{\alpha}^{\dagger}(\rv_1) \psi_{\alpha}^{\phantom{\dagger}}(\rv_1)
       \psi_{\beta}^{\dagger}(\rv_2) \psi_{\beta}^{\phantom{\dagger}}(\rv_2) \ . \nonumber
\end{eqnarray}
The operator $\psi^\dagger_\alpha(\rv)$ creates a particle of spin $\alpha\in\lbrace\uparrow,\downarrow\rbrace$ at position $\rv$, and the summation over repeated spin indices is implicit. The lattice $V(\rv)$ gives rise to a band structure of energy levels, and the density of particles is tuned to any number of completely populated bands at zero temperature by placing the chemical potential $\mu$ in a bandgap. We shall describe the dynamics of the resulting band insulator by an effective theory of low-energy quasiparticles belonging to the valence and conduction bands. Of particular interest will be sufficiently strong attractive interactions $U$ which can drive the system into a superfluid state. A similar instability can be created by bringing the chemical potential sufficiently close to a band edge, even at weak couplings. A qualitative example of the superfluid-insulator transitions is shown in Fig.\ref{pd1}.

The quadratic part of the Hamiltonian (\ref{ContModel1}) can be diagonalized by switching to the band representation:
\begin{equation}
H_0 = \sum_n \int \frac{\dd^d k}{(2\pi)^d} \psi_{n,\kv,\alpha}^\dagger \left\lbrack \varepsilon_n(\kv) - \mu
   \right\rbrack \psi_{n,\kv,\alpha}^{\phantom{\dagger}} \ ,
\end{equation}
where the operator $\psi^\dagger_{n, \kv, \alpha}$ creates a particle in the band $n$ with the crystal momentum $\kv$. The band structure energy levels are $\varepsilon(\kv)$, and crystal momenta are integrated out in the first Brillouin zone. This representation change also replaces the spatial dependence of interactions with the dependence on crystal momenta $\kv_1$, $\kv_2$ of the incoming particles, as well as the momentum transfer $\qv$ in the collision. Since two fermions from any pair of bands $(n_1,n_2)$ can scatter into any other pair of bands $(m_1,m_2)$, we need to keep track of all interaction channels governed by couplings $\Wt{U}_{n_1 n_2}^{m_1 m_2}(\kv_1,\kv_2,\qv)$ in the band representation. For example, a pure contact interaction $U \psi_{\alpha}^{\dagger}(\rv) \psi_{\alpha}^{\phantom{\dagger}}(\rv) \psi_{\beta}^{\dagger}(\rv) \psi_{\beta}^{\phantom{\dagger}}(\rv)$ would produce:
\begin{eqnarray}\label{BandInteractions}
&& \!\!\! \Wt{U}_{n_1 n_2}^{m_1 m_2}(\kv_1,\kv_2,\qv) = \\
&& ~~ U \int\limits_{\Tr{UC}} \dd^d r \; u_{m_1,\kv_1+\qv}^*(\rv) u_{n_1,\kv_1}^{\phantom{*}}(\rv)
  u_{m_2,\kv_2-\qv}^*(\rv) u_{n_2,\kv_2}^{\phantom{*}}(\rv) \nonumber
\end{eqnarray}
where UC indicates the spatial integration over the lattice unit-cell, and $u_{n,\kv}(\rv)$ originate from the Bloch wave-functions $\Wt{\psi}_{n,\kv}(\rv) = u_{n,\kv}(\rv) e^{i\kv\rv}$. This expression illustrates an important property of interactions in the band representation which follows from the overlap features of the Bloch wavefunctions. As a rule of thumb, the couplings $\Wt{U}_{n_1 n_2}^{m_1 m_2}$ are largest by magnitude if $n_i=m_i$ for both $i=1,2$ and smallest if $n_i \neq m_i$ for both $i=1,2$. The strongest interaction channels involve a single band, while the interband couplings are weaker. This is a natural situation for generic band structures and short-range interactions, but it could be reversed in principle.

Before turning to the functional formalism, we must normal-order the interaction part of the Hamiltonian:
\begin{eqnarray}
H_{\Tr{int}} &=& \sum_{n_1 m_1} \sum_{n_2 m_2}
      \int \frac{\dd^d k_1}{(2\pi)^d} \frac{\dd^d k_2}{(2\pi)^d} \frac{\dd^d q}{(2\pi)^d}
        \; \Wt{U}_{n_1 n_2}^{m_1 m_2}(\kv_1,\kv_2,\qv) \nonumber \\
&& ~~~~~ \times \psi_{m_1,\kv_1+\qv,\alpha}^{\dagger} \psi_{n_1,\kv_1,\alpha}^{\phantom{\dagger}}
      \psi_{m_2,\kv_2-\qv,\beta}^{\dagger} \psi_{n_2,\kv_2,\beta}^{\phantom{\dagger}} \nonumber \\
&=& \sum_{n_1 m_1} \sum_{n_2 m_2}
      \int \frac{\dd^d k_1}{(2\pi)^d} \frac{\dd^d k_2}{(2\pi)^d} \frac{\dd^d q}{(2\pi)^d}
        \; \Wt{U}_{n_1 n_2}^{m_1 m_2}(\kv_1,\kv_2,\qv) \nonumber \\
&& ~~~~~ \times \psi_{m_1,\kv_1+\qv,\alpha}^{\dagger} \psi_{m_2,\kv_2-\qv,\beta}^{\dagger}
      \psi_{n_2,\kv_2,\beta}^{\phantom{\dagger}} \psi_{n_1,\kv_1,\alpha}^{\phantom{\dagger}} \nonumber \\
&&  + \sum_{nm} \int \frac{\dd^d k}{(2\pi)^d} \; \Wt{U}_{n}^{m}(\kv)
      \psi_{m,\kv,\alpha}^{\dagger} \psi_{n,\kv,\alpha}^{\phantom{\dagger}}
\end{eqnarray}
This generates quadratic terms
\begin{equation}
\Wt{U}_{n}^{m}(\kv) = \sum_{n'} \int \frac{\dd^d q}{(2\pi)^d} \Wt{U}_{n' n}^{m n'}(\kv-\qv,\kv,\qv) \sim
   \frac{n_0}{2}\sum_{n'} \Wt{U}_{n' n}^{m n'} \ . \nonumber
\end{equation}
For interactions which do not extend beyond a single unit-cell in real space, $\Wt{U}_{n_1 n_2}^{m_1 m_2}(\kv_1,\kv_2,\qv) \approx \Wt{U}_{n_1 n_2}^{m_1 m_2}$ so $\Wt{U}_{n}^{m}(\kv)$ are proportional to the particle density $n_0$ in the ground-state. The diagonal $\Wt{U}_n^n$ are ``charging energies'' which shift the chemical potential and quasiparticle gaps due to interactions.

In order to derive the effective theory we convert the Hamiltonian $H_0$ + $H_{\Tr{int}}$ into an imaginary-time action and formally integrate out the fermion fields from high-energy bands in the obtained path integral. At best, this can be done perturbatively, for example using the Feynman diagram technique. The action retains only a few low-energy fields, corresponding to quasiparticles and quasiholes which may live in the conduction and valence bands at multiple symmetry-related wavevectors in the first Brillouin zone. Additional quadratic band-mixing couplings of the $U_n^m$ kind arise from virtual scattering between low energy bands via high energy states. However, at this point we re-diagonalize the quadratic terms by another change of representation and obtain the following generic form of the effective theory in which all band-mixing quadratic terms are eliminated:
\begin{eqnarray}\label{Seff}
&& S_{\Tr{eff}} = \sum_n \int \frac{\dd\omega}{2\pi} \frac{\dd^d k}{(2\pi)^d} \;
          f_{n,k,\alpha}^{\dagger} \left\lbrack -i\omega + E_n(\kv) \right\rbrack f_{n,k,\alpha}^{\phantom{\dagger}}
  \nonumber \\ && ~~ +
    \sum_{n_1 m_1} \sum_{n_2 m_2} U_{n_1 n_2}^{m_1 m_2} \int
       \frac{\dd\omega_1}{2\pi} \frac{\dd^d k_1}{(2\pi)^d}
       \frac{\dd\omega_2}{2\pi} \frac{\dd^d k_2}{(2\pi)^d}
       \frac{\dd\Omega}{2\pi} \frac{\dd^d q}{(2\pi)^d}
     \nonumber
  \\ & & ~~~~ \times f_{m_1,k_1+q,\alpha}^{\dagger} f_{m_2,k_2-q,\beta}^{\dagger}
                     f_{n_2,k_2,\beta}^{\phantom{\dagger}} f_{n_1,k_1,\alpha}^{\phantom{\dagger}} \ .
\end{eqnarray}
We used the notation $k_i=(\omega_i,\kv_i)$ and $q=(\Omega,\qv)$. The Grassmann fields $f_{n,k,\alpha}$ now correspond to quasiparticles whose flavor is a perturbative mixture of different band fermions, but retains a dominant (unperturbed) conduction or valence band contribution due to the bandgap in which the chemical potential sits. In this sense the bandgap protects the fundamental character of the quantum numbers $n$, which label all low energy quasiparticle species. All interaction couplings $U$ are effective, as well as the gapped quasiparticle dispersions $E_n(\kv)$.

The goal of this paper is to explore the universal properties of nearly resonantly interacting fermions in periodic potentials, for the purpose of which we apply renormalization group. Therefore, the precise functional forms and values of the effective $E_n(\kv)$ and $U_{n_1 n_2}^{m_1 m_2}(\kv_1,\kv_2,\qv)$ are not of interest, and we do not attempt to derive them from any microscopic model such as (\ref{ContModel1}). We will simply consider the most generic forms which affect the universal regimes of band insulators and nearby states. These forms are $E_n(\kv)=E_{n0}+k^2/2m_n$ with effective gaps $E_{n0}$ and masses $m_n$, and $U_{n_1 n_2}^{m_1 m_2}$ without any crystal momentum dependence as written in (\ref{Seff}). No other terms allowed by symmetries are important for the universal physics near the RG fixed points. Furthermore, we will consider in greater detail only the cases in which the indices $n_i, m_i$ denote one or two bands. The most generic transitions are driven by the chemical potential $\mu$, so that only one band is important (see Fig.\ref{pd1}, (p) and (h) trajectories). The transitions involving particles and holes (the (ph) trajectory in Fig.\ref{pd1}) require at least two bands for a complete description.

A potential problem recognized in a number of studies is that in the vicinity of resonant scattering the microscopic interactions may correspond to energy scales (much) larger than the bandgap. Then, many high energy bands may be significantly hybridized with the conduction and valence bands. In the present formulation of the problem, this can lead to a strong renormalization of low-energy quasiparticle dispersions $E_n(\kv)$ and effective interactions $U_{n_1 n_2}^{m_1 m_2}$. In most circumstances this does not endanger the analysis which follows.

As a result of integrating out high energy bands, $E_n(\kv)$ and $U_{n_1 n_2}^{m_1 m_2}$ are perturbative expansions in powers of the ratio $U/E_{he}$, where $U$ is the microscopic interaction strength and $E_{he}$ is the smallest energy of the integrated high-energy fermions. These expansions need not converge fast, but they must be convergent, otherwise the effective $E_n(\kv)$ and $U_{n_1 n_2}^{m_1 m_2}$ would contain singular features which would invalidate our analysis. In general there should be a minimum number of bands which must be kept in the effective theory in order for it to be properly analytic, and this number may grow when one approaches the resonant scattering in empty space at $V(\rv)\equiv 0$. We shall assume that this number is never larger than two in the present cases of interest. An assumption of this kind is made in all other studies and it is justified by the fact that a lattice potential shifts the scattering resonance from its empty-space position toward the effective lattice BEC limit \cite{Fedichev2004, Koetsier2006}. In other words, the effective unitarity limit in a lattice which we seek to describe corresponds to the BCS regime in empty space for the same microscopic interactions, where $U/E_{he}$ is not too large.

\section{Renormalization group analysis}\label{secRG}

Here we apply renormalization group (RG) to band insulators in the unitarity regime. We will identify various fixed points associated with unitarity (resonant two-body scattering) which emerge in the presence of an external periodic potential $V(\rv)$, but otherwise are analogous to the unitarity fixed point of a uniform system at $V(\rv)\equiv 0$. The main difference is that the fixed points we shall discuss occur at finite densities of microscopic particles, corresponding to fully occupied bands, whereas universality in the uniform system stems from a \emph{zero density} fixed point.

There are two characteristic situations which will be considered separately. First, one fermion species (either particles or holes) generally dominates dynamics in the unitarity regime, so the renormalization group equations can be derived exactly to all orders of perturbation theory. This is extremely useful because run-away flows of interaction couplings can be traced more reliably. The second situation is more special and occurs when both particles and holes participate equally in dynamics. Then, the fixed point structure becomes intricate, but can be accessed only in an $\epsilon$ (or large-$N$) expansion. At the end we briefly discuss extensions to more realistic cases with multiple relevant fermion species, and measurable manifestations of the universality class.

\subsection{Transitions involving one fermion species}\label{secP}

A transition dominated by either particles or holes, but not both, is generally caused by chemical potential changes as illustrated in Fig.\ref{pd1} with (p) and (h) dashed lines. As a natural starting point one can imagine a band insulator either in the deep BCS limit, or with a very deep lattice potential, where the chemical potential is brought much closer to one of the bands than to the other. However, such extreme regimes are not necessary initially because they are created by the RG flow (the bandgap is a relevant operator). We shall discover that an attractive interaction undergoes a run-away flow in two dimensions and competes with the flow of bandgap. This competition is resolved at cut-off scales where the RG breaks down. However, a new strongly-coupled universality class takes over in that limit, associated with superfluid to Mott-insulator transition with dynamical exponent $z=2$ at the intersection of the thick red and dashed-blue lines (p) or (h) in Fig.\ref{pd1}.

A characteristic weak-coupling fixed point contained in a theory of attractively interacting fermions is unitarity. It is found at zero temperature when the strength of interactions is tuned to a critical value and the chemical potential lies exactly at the boundary between a fermion band and a bandgap (or vacuum). This fixed point is characterized by its own ``unitarity'' universality class \cite{Nishida2007, nikolic:033608} and corresponds to a special zero-temperature pairing instability of band-insulators, which is different than the standard BCS instability of metals at finite temperatures.

The simplest possible effective action (\ref{Seff}) of a band insulator contains a single species of interacting spinful fermions which live in the band closest to the chemical potential. These fermions are either particle excitations from the conduction band, or hole excitations from the valence band. We assume that either kind of excitations is gapped and quadratically dispersing at the lowest energies near a single point in the first Brillouin zone. By scaling, only the zero-range part of interactions and the chemical potential are relevant operators at $T=0$ near the unitarity fixed point, so the effective theory in $d$ dimensions is the same as for free fermions in the absence of the lattice:
\begin{eqnarray}\label{CritTheory1}
S_1 & = & \int \mathcal{D}k \; f_{k,\alpha}^{\dagger} \left\lbrack -i\omega + E(\kv) \right\rbrack
    f_{k,\alpha}^{\phantom{\dagger}} \\ & + &
    U \int \mathcal{D}k_1 \mathcal{D}k_2 \mathcal{D}q \;
      f_{k_1+q,\alpha}^{\dagger} f_{k_2-q,\beta}^{\dagger}
      f_{k_2,\beta}^{\phantom{\dagger}} f_{k_1,\alpha}^{\phantom{\dagger}}
  \nonumber \ ,
\end{eqnarray}
where $k=(\omega,\kv)$, $\mathcal{D}k = \dd \omega \dd^d \kv / (2\pi)^{d+1}$, and
\begin{equation}
E(\kv) = E_0 + \frac{k^2}{2m} \ .
\end{equation}
Here, $E_0$ is the bare gap of fermionic excitations, and $m$ is their effective mass.

\begin{figure}
\includegraphics[width=2.2in]{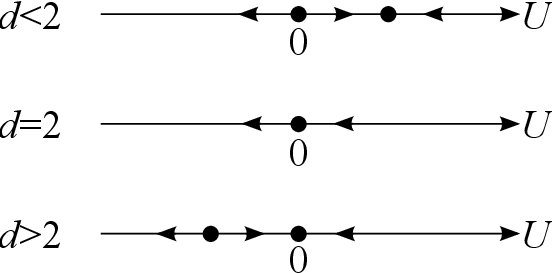}
\caption{\label{RG1}The RG flow of the interaction coupling $U$ in the theory (\ref{CritTheory1}).}
\end{figure}

We will now examine the effect of interactions on a $T=0$ band insulator by applying the perturbative RG to the above theory. Since the ground state is always a vacuum of quasiparticles, all Feynman diagrams which contain fermion loops must vanish (formally, the unequal scaling of space and time in our non-relativistic theory is handled by integrating all frequencies in the coarse-graining step of RG; these frequency integrals vanish since the poles of all Green's functions on a loop lie in the same complex half-plane). As a consequence, only the interaction coupling is renormalized by a summable geometric progression of ladder diagrams, which do not contain fermion loops. The exact RG equations under a rescaling of length scales by a factor of $e^l$ are found to be \cite{nikolic:033608, SubirQPT}:
\begin{equation}\label{RGflow1}
\frac{\dd E_0}{\dd l} = 2 E_0 \qquad , \qquad \frac{\dd U}{\dd l} = (2-d)U - \Pi U^2 \ ,
\end{equation}
where $l$ is the scale parameter, and $\Pi$ is a positive cutoff-dependent constant derived in appendix \ref{app1} (strictly speaking, $\Pi$ decreases with the running $E_0$, but this can be neglected while $E_0$ is much smaller than the cut-off energy). A fixed point is always found at $E_0=0$, $U=0$. An additional non-trivial fixed point is found in $d\neq 2$ at $E_0=0$, $U=U^*=(2-d)\Pi^{-1}$, which describes attractive interactions if $d>2$ and corresponds to unitarity. The schematic flow of interaction couplings is shown in Fig.\ref{RG1}.

Any attractive interaction $U<0$ in $d=2$ has a run-away flow to $U\to-\infty$, while in $d>2$ it needs to be large enough to flow toward $U\to-\infty$. Repulsive interactions $U>0$, on the other hand, flow to the Gaussian fixed point in $d\ge 2$. Therefore, only attractive interactions can produce strongly correlated states. Since the RG equations are exact, we can precisely characterize the run-away flow, assuming that energy/momentum dependence of interactions, and the effects of multi-body collisions remain negligible at least in the insulating state (see appendix \ref{app1} for the justification of this assumption). Solving for $U(l)$ in two and three dimensions we obtain:
\begin{equation}\label{RGflow2}
  U(l) = \left\lbrace
    \begin{array}{lcl}
      \frac{U(0)}{1+\Pi U(0)l} & , & d=2 \\[0.1in]
      \frac{U(0)}{\lbrack 1+\Pi U(0) \rbrack e^l-\Pi U(0)} & , & d=3
    \end{array}
  \right\rbrace
\end{equation}
In both cases, the run-away flows have vertical asymptotes so that $U(l)$ diverges at a finite value of $l$ ($l=|\Pi U(0)|^{-1}$ in $d=2$). This indicates that Cooper pairs become stable at a finite length scale and have a finite coherence length despite the fermion bandgap. However, the interpretation of the run-away flow breaks-down at the cut-off scale because extremely large $U$ will pair-up the high-energy fermions, which were assumed to be unpaired in this RG procedure. A boson-dominated dynamics takes over the shortest length-scales under consideration.

Note that $U(l)\to-\infty$ at a finite $l$ cannot be immediately interpreted as a signal of superfluidity. This is due to the fact that at finite $l$ we do not yet have a theory which transparently describes dynamics at macroscopic scales, while superfluidity is verified only by macroscopic long-range correlations. Since RG is based on integrating out \emph{high energy} modes, it does not provide a precise answer to the question of what phase the system lives in, but only gives an indication.

The fermion gap in an insulating state grows exponentially under RG, $E_0(l) = E_0(0) e^{2l}$. If $E_0(l)$ is the first to reach the cutoff energy scale, the further RG flow is halted (RG breaks down) in a state apparently devoid of particles. This is a band insulator. Note that in this case the run-away flows of interactions in (\ref{RGflow2}) are gradually slowed down and eventually halted by the dependence of $\Pi$ on the running $E_0$ (detailed in the appendix \ref{app1}). A band insulator is obtained in $2+\epsilon$ dimensions for any sufficiently weak interaction $|U(0)|<|U^*|\sim\epsilon$, or even for $|U(0)|>|U^*|$ provided that the microscopic gap $E_0(0)$ is large enough.

If $U(l)$ is the first to reach its cutoff instead, then boson-dominated dynamics at shortest length-scales requires switching to a purely bosonic effective theory in order to determine what happens at large length-scales. Both insulating and superfluid phases are possible in this limit, despite a finite fermion gap, but the transition between them is in a different universality class than the BCS pair-breaking transition.

\subsection{Transitions involving particles and holes}\label{secPH}

A special case is obtained in the vicinity of vanishing gaps for both particle and hole excitations. A pairing transition influenced by a corresponding fixed point cannot be obtained by changing the chemical potential alone, but can be accessed by tuning interaction strength or lattice depth, at a fixed particle density (see Fig.\ref{pd1}, the (ph) dashed line). The values of $\mu$ and $V$ should lie at the intersection of effective conduction and valence bands where the effective bandgap closes. We shall again discover a run-away flow of interaction couplings, but this time it quickly invalidates the perturbative RG. The run-away flow is expected to eventually lead to a strong-coupling fixed point in the XY universality class, associated with the superconductor to Mott-insulator transition at an integer number of bosons per lattice site.

Assuming that the bandgap is direct and smallest at the $\Gamma$ point of the first Brillouin zone, we write the critical theory (\ref{Seff}) for valence ($\Tr{v}$) and conduction ($\Tr{c}$) electrons:
\begin{eqnarray}\label{S2}
S_2 & = & \sum_n \int \mathcal{D}k \;
          f_{n,k,\alpha}^{\dagger} \left\lbrack -i\omega + E_n(\kv) \right\rbrack f_{n,k,\alpha}^{\phantom{\dagger}}
  \\ & + &
    \sum_{n_1 m_1} \sum_{n_2 m_2} U_{n_1 n_2}^{m_1 m_2} \int \mathcal{D}k_1 \mathcal{D}k_2 \mathcal{D}q
     \nonumber
  \\ & & ~~ \times f_{m_1,k_1+q,\alpha}^{\dagger} f_{m_2,k_2-q,\beta}^{\dagger}
                   f_{n_2,k_2,\beta}^{\phantom{\dagger}} f_{n_1,k_1,\alpha}^{\phantom{\dagger}}
  \nonumber \ ,
\end{eqnarray}
where $n_i,m_i\in\lbrace \Tr{c},\Tr{v} \rbrace$ and
\begin{equation}
E_{\Tr{v}}(\kv) = -E_{\Tr{v}0} - \frac{k^2}{2m_{\Tr{v}}} \quad , \quad
E_{\Tr{c}}(\kv) = E_{\Tr{c}0} + \frac{k^2}{2m_{\Tr{c}}} \nonumber \ .
\end{equation}
Here, $E_{\Tr{c}0}$ and $E_{\Tr{v}0}$ are excitation gaps, and $m_{\Tr{c}}$ and $m_{\Tr{v}}$ are effective masses of particle and hole excitations respectively. Other circumstances with indirect bandgaps and multiple fermion species living near different wavevectors in the Brillouin zone will be discussed in the following section. Performing a particle-hole transformation for the valence band cannot help us construct an exact RG procedure. Instead, it is convenient to work directly with the native particle degrees of freedom.

It is worth noting that the spatial dependence of the microscopic interaction potential $U(\rv)$ on the distance $\rv$ between the interacting particles is not automatically irrelevant (in the RG sense) in the presence of the lattice. Short-range variations of $U(\rv)$, at or below the lattice spacing length $a_L$, affect the relative strength of the couplings $U_{n_1 n_2}^{m_1 m_2}$, which may lead to non-trivial interacting fixed points as discussed below. Only the variations of $U(\rv)$ at $r \gtrsim a_L$ scale to zero under RG in the vicinity of the fixed points of interest, allowing us to discard the ensuing crystal momentum dependence of $U_{n_1 n_2}^{m_1 m_2}$ as irrelevant.

Even though the effective action (\ref{S2}) contains only diagonal quadratic terms, off-diagonal quadratic couplings $U_n^m$ are generated by RG if interactions do not conserve the parity of the particle number in individual bands. We shall see later why this happens at larger length scales, but for now we need to devise a formalism to deal with such running band-mixing couplings. At least we can guarantee that the generated $U_n^m$ will never have momentum dependence, because mass scales are kept fixed under RG and the original effective action does not contain any band-mixing mass terms $\sim k^2/2m$.

The additional band-mixing terms $U_n^m f_{m,k,\alpha}^\dagger f_{n,k,\alpha}^{\phantom{\dagger}}$ in the action affect the bare quasiparticle Green's function. One way to derive this Green's function is to treat $U_{n}^{m}$ as a self-energy correction to the purely diagonal non-interacting Green's function $\lbrack i\omega-E_n(\kv) \rbrack^{-1} \delta_{nm}$:
\begin{eqnarray}
G_n^m(\kv,i\omega) & = & \left(
  \begin{array}{cc}
     i\omega-E_{\Tr{c}}(\kv)-U_{\Tr{c}}^{\Tr{c}} & -U_{\Tr{c}}^{\Tr{v}} \\
     -U_{\Tr{v}}^{\Tr{c}} & i\omega-E_{\Tr{v}}(\kv)-U_{\Tr{v}}^{\Tr{v}}
  \end{array}
\right)^{-1} \nonumber \\
& = & \frac{g_n^m(\kv,i\omega)}{(i\omega-z_1(\kv))(i\omega-z_2(\kv))} \ .
\end{eqnarray}
It is convenient to define $\zeta_n(\kv) = E_n(\kv) + U_n^n$ and $\xi = \sqrt{U_{\Tr{c}}^{\Tr{v}}U_{\Tr{v}}^{\Tr{c}}}$ (note that $U_{\Tr{c}}^{\Tr{v}} = (U_{\Tr{v}}^{\Tr{c}})^*$). Then:
\begin{equation}
z_{1/2}(\kv) = \frac{\zeta_{\Tr{c}}+\zeta_{\Tr{v}}}{2} \pm \left\lbrack
  \left( \frac{\zeta_{\Tr{c}}-\zeta_{\Tr{v}}}{2} \right)^2 + \xi^2 \right\rbrack^{\frac{1}{2}}
\end{equation}
are the new poles of the bare fermion excitations, and
\begin{equation}
g_n^m(\kv,i\omega) = (i\omega - \zeta_{\Tr{c}} - \zeta_{\Tr{v}} + \zeta_n) \delta_{nm} + \xi (1-\delta_{nm}) \ .
\end{equation}
It is easy to show that both poles are always real, one being positive (particle-like) and the other negative (hole-like) as long as $\xi^2>\zeta_{\Tr{c}}\zeta_{\Tr{v}}$. We will assume that the latter condition is satisfied, so that the system is a band insulator. Consequently, we can expand the poles up to the highest relevant order $\mathcal{O}(k^2)$:
\begin{equation}
z_1(\kv) = \epsilon_1 + \frac{k^2}{2M_1} \qquad , \qquad
z_2(\kv) = -\epsilon_2 - \frac{k^2}{2M_2} \ ,
\end{equation}
where $\epsilon_i$ are the bare quasiparticle gaps, and $M_i$ are the quasiparticle masses given by:
\begin{equation}
M_{1/2}^{-1} = \alpha \times \frac{m_{\Tr{c}}^{-1} +
   m_{\Tr{v}}^{-1}}{2} \pm \frac{m_{\Tr{c}}^{-1} - m_{\Tr{v}}^{-1}}{2} \ .
\end{equation}
The parameter
\begin{equation}\label{PrmAlpha}
\alpha = \frac{E_g}{\sqrt{E_g^2+4\xi^2}}
\end{equation}
captures the amount of mixing between the two bands ($0 \le \alpha \le 1$); $E_g = ( \zeta_{\Tr{c}}-\zeta_{\Tr{v}}) \bigr\vert_{\kv=0}$ is the effective fermion bandgap. For $\alpha=1$ there is no band mixing since $M_i \in \lbrace m_{\Tr{c}}, m_{\Tr{v}} \rbrace$. In general, $\alpha > |\beta| = |m_{\Tr{c}}-m_{\Tr{v}}|/(m_{\Tr{c}}+m_{\Tr{v}})$ is required in order for both $M_i$ to remain positive. Otherwise, band inversion is caused by large interband couplings and it must be taken into account by redefining the low-energy quasiparticles, which then live at some different momenta in the first Brillouin zone. We shall come back to this situation at the end.

Now we set up the RG. As usual, we keep the masses $m_{\Tr{c}}$ and $m_{\Tr{v}}$ in the absence of interactions fixed under RG. While this does not imply that $M_i$ will be fixed, it sets the scaling dimension for the field operators to $d/2$. The scaling of coordinates and couplings
\begin{eqnarray}
r' = r e^{-l} \qquad & , & \qquad \tau' = \tau e^{-2l} \\
\epsilon'_i = \epsilon_i e^{2l} \qquad  & , & \qquad U' = U e^{(2-d)l} \ , \nonumber
\end{eqnarray}
is followed by the diagrammatic integration of high-energy fields living at all Matsubara frequencies and momenta within a shell $|\kv|\in(\Lambda e^{-\dd l}, \Lambda)$, where $\Lambda$ is a cut-off momentum scale and $\dd l$ is an infinitesimal increment of the scale parameter $l$. The resulting one-loop renormalization of the quadratic and quartic couplings is summarized in table \ref{OneLoopRen}. The relevant cutoff-dependent renormalization scales are:
\begin{eqnarray}\label{RenConst}
K_{1k} & = & \frac{S_d \Lambda^d}{(2\pi)^d} \frac{\alpha-(-1)^k}{2\alpha} \\
K'_{2kk'} & = & \frac{S_d \Lambda^{d-2}}{(2\pi)^d}
  \frac{-m_{\Tr{c}}m_{\Tr{v}} (1+\alpha^2-2\delta_{kk'})}{2\alpha^3(m_{\Tr{c}}+m_{\Tr{v}})} \nonumber \\
K''_{2kk'} & = & \frac{S_d \Lambda^{d-2}}{(2\pi)^d}
  \frac{m_{\Tr{c}}m_{\Tr{v}} \left\lbrack (m_{\Tr{c}}+m_{\Tr{v}})(\alpha^2-1) + 4m_k\delta_{kk'} \right\rbrack}
       {2\alpha \left\lbrack \alpha^2( m_{\Tr{c}}+m_{\Tr{v}})^2 - (m_{\Tr{c}}-m_{\Tr{v}})^2 \right\rbrack} \nonumber
\end{eqnarray}
where $S_d=2\pi^{d/2}/\Gamma(d/2)$ is the $d$-dimensional unit-sphere area.

\begin{table}[!]
  \begin{displaymath}
    \begin{array}{cc}
      \begin{minipage}{1.2in} \includegraphics[width=0.7in]{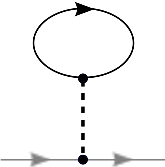} \end{minipage} &
        -2\sum\limits_k  K_{1k}^{\phantom{mk}} U_{nk}^{mk} \\[0.5in]
      \begin{minipage}{1.2in} \includegraphics[width=1.2in]{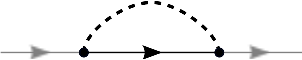} \end{minipage} &
        \sum\limits_k K_{1k}^{\phantom{mk}} U_{kn}^{mk} \\[0.5in]
      \begin{minipage}{1.2in} \includegraphics[width=1.2in]{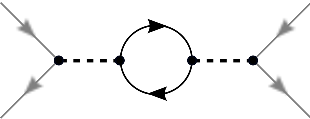} \end{minipage} &
        \begin{array}{r}
          -2 \sum\limits_{kl} K'_{2kl}
          \Bigl( U_{n_1 k}^{m_1 l} U_{n_2 l}^{m_2 k} + U_{n_1 k}^{m_1 l} U_{l n_2}^{k m_2} \\
               + U_{k n_1}^{l m_1} U_{n_2 l}^{m_2 k} + U_{k n_1}^{l m_1} U_{l n_2}^{k m_2} \Bigr)
        \end{array} \\[0.5in]
      \begin{minipage}{1.2in} \includegraphics[width=1.0in]{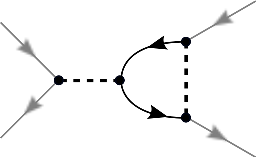} \end{minipage} &
        \begin{array}{r}
          \sum\limits_{kl} K'_{2kl}
          \Bigl( U_{n_1 k}^{m_1 l} U_{n_2 l}^{k m_2} + U_{n_1 k}^{m_1 l} U_{l n_2}^{m_2 k} \phantom{\Bigr)} \\
               + U_{k n_1}^{l m_1} U_{n_2 l}^{k m_2} + U_{k n_1}^{l m_1} U_{l n_2}^{m_2 k} \phantom{\Bigr)} \\
               + U_{n_2 k}^{m_2 l} U_{n_1 l}^{k m_1} + U_{n_2 k}^{m_2 l} U_{l n_1}^{m_1 k} \phantom{\Bigr)} \\
               + U_{k n_2}^{l m_2} U_{n_1 l}^{k m_1} + U_{k n_2}^{l m_2} U_{l n_1}^{m_1 k} \Bigr)
        \end{array} \\[0.5in]
      \begin{minipage}{1.2in} \includegraphics[width=0.85in]{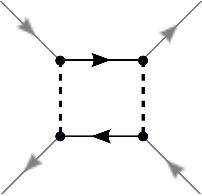} \end{minipage} &
        \begin{array}{r}
          \sum\limits_{kl} K'_{2kl}
          \Bigl( U_{k n_2}^{m_1 l} U_{n_1 l}^{k m_2} + U_{k n_2}^{m_1 l} U_{l n_1}^{m_2 k} \phantom{\Bigr)} \\
               + U_{n_2 k}^{l m_1} U_{n_1 l}^{k m_2} + U_{n_2 k}^{l m_1} U_{l n_1}^{m_2 k} \Bigr)
        \end{array} \\[0.5in]
      \begin{minipage}{1.2in} \includegraphics[width=0.85in]{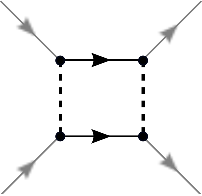} \end{minipage} &
        \begin{array}{r}
          \sum\limits_{kl} K''_{2kl}
          \Bigl( U_{kl}^{m_1 m_2} U_{n_1 n_2}^{kl} + U_{kl}^{m_1 m_2} U_{n_2 n_1}^{lk} \\
               + U_{lk}^{m_2 m_1} U_{n_1 n_2}^{kl} + U_{lk}^{m_2 m_1} U_{n_2 n_1}^{lk} \Bigr)
        \end{array}
    \end{array}
  \end{displaymath}
\caption{\label{OneLoopRen}One-loop diagrams which renormalize the couplings $U_n^m$ (first two) and $U_{n_1 n_2}^{m_1 m_2} $ (last four). The renormalization constants $K_{1k}$, $K'_{2kk'}$ and $K''_{2kk'}$ are given in (\ref{RenConst}).}
\end{table}

The RG equations involving all four $U_n^m$ and all sixteen $U_{n_1 n_2}^{m_1 m_2}$ couplings (not all of which are independent) are too complicated to be fully solved. A part of the problem is that the parameter $\alpha$ can also flow under RG, as a result of the renormalization of the couplings $U_n^m$. In order to simplify notation let us combine the bare fermion gaps and any generated diagonal couplings $U_n^n$ into the same symbol: $U_n^n + E_{n0} \to U_n^n$. We begin by noting that the RG equation for the quadratic couplings is:
\begin{equation}\label{RGquad}
~~ \frac{\dd U_n^m}{\dd l} = 2 U_n^m - \sum_k K_{1k} \left( U_{kn}^{mk} + U_{nk}^{km} - 2U_{nk}^{mk} - 2U_{kn}^{km} \right)
 \ .
\end{equation}
In $d=2+\epsilon$ dimensions the interacting fixed points will be at $U_{n_1n_2}^{m_1m_2} \propto \epsilon$, implying $U_n^m \propto \epsilon$. Finite values for all $U_n^m$ in $d>2$ dimensions uniquely determine the value for $\alpha$, which has to be fed back into the RG equations to self-consistently determine the fixed points. This can be done only numerically. However, analytical solutions for a subset of fixed points can be found if the couplings $U_{\Tr{c}}^{\Tr{v}}$, $U_{\Tr{cc}}^{\Tr{cv}}$, $U_{\Tr{cv}}^{\Tr{cc}}$, $U_{\Tr{vv}}^{\Tr{cv}}$, $U_{\Tr{cv}}^{\Tr{vv}}$ and their complex conjugates are all zero. These interactions do not conserve the parity of the particle numbers in individual bands, but if all of them are zero then it follows from (\ref{PrmAlpha}) and (\ref{RGquad}) that $\alpha=1$ and does not flow under RG. This will be the focus of the following discussion.

In two dimensions there is only one weak-coupling fixed point, at $U_n^m=0$, $U_{n_1 n_2}^{m_1 m_2}=0$. The flow of $U_\Tr{cc}^\Tr{cc}$, $U_\Tr{vv}^\Tr{vv}$ and $U_\Tr{cv}^\Tr{vc}=U_\Tr{vc}^\Tr{cv}$ is of the same type as shown in Fig.\ref{RG1} at $d=2$, so that the attractive interactions undergo a run-away flow, while repulsive interactions flow to zero. The interband interaction $U_\Tr{cv}^\Tr{cv}=U_\Tr{vc}^\Tr{vc}$ has the opposite behavior, a repulsive one keeps growing, while an attractive one flows to zero. In normal circumstances, due to the Bloch wavefunction properties, the intraband couplings $U_\Tr{cc}^\Tr{cc}$ and $U_\Tr{vv}^\Tr{vv}$ are larger than the interband $U_\Tr{cv}^\Tr{cv}$ and $U_\Tr{cv}^\Tr{vc}$, so the attractive intraband channels dominate at macroscopic scales and lead to Cooper pairing even if the interband channels are repulsive. Instabilities in the particle-hole channel are possible only if all interactions are repulsive, or if for some reason $U_\Tr{cv}^\Tr{cv}$ is repulsive and stronger than the attractive intraband interactions.

\begin{table}[t]
   \begin{tabular}{c@{\;\;\;\;\;}c@{\;\;\;\;\;}c}
      \includegraphics[height=0.45in]{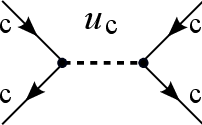} &
      \includegraphics[height=0.45in]{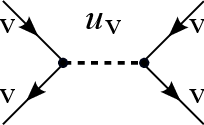} &
      \includegraphics[height=0.45in]{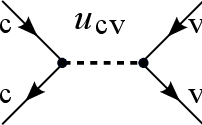} \\[0.2in]
      \includegraphics[height=0.45in]{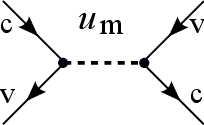} &
      \includegraphics[height=0.45in]{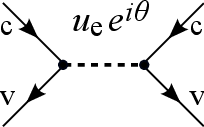} &
      \includegraphics[height=0.45in]{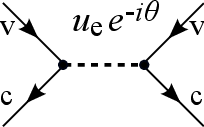}
   \end{tabular}
\caption{\label{TabVertices}Relevant interaction vertices near the zero-quasiparticle-density fixed points for $\alpha=1$ (defined in the text).}
\end{table}

In $d=2+\epsilon$ dimensions with $\epsilon>0$ it is convenient to define:
\begin{equation}
m_{\Tr{c}} = m(1+\beta) \quad , \quad m_{\Tr{v}} = m(1-\beta) \quad , \quad
  \beta = \frac{m_{\Tr{c}}-m_{\Tr{v}}}{m_{\Tr{c}}+m_{\Tr{v}}} \nonumber
\end{equation}
and the rescaled independent dimensionless couplings $(u_{\Tr{c}}, u_{\Tr{v}}, u_{\Tr{cv}}, u_{\Tr{m}}, u_{\Tr{e}}, e_{\Tr{g}})$:
\begin{eqnarray}\label{ScalInt}
U_{\Tr{c}}^{\Tr{c}} - U_{\Tr{v}}^{\Tr{v}} & = & \frac{\Lambda^2\epsilon}{m} e_g
\\
  U_{\Tr{cc}}^{\Tr{cc}} = K\epsilon \frac{u_{\Tr{c}}}{1+\beta} ~ &,& ~
  U_{\Tr{vv}}^{\Tr{vv}} = K\epsilon \frac{u_{\Tr{v}}}{1-\beta}
\nonumber \\
  U_{\Tr{cv}}^{\Tr{cv}} + U_{\Tr{vc}}^{\Tr{vc}} = K\epsilon \frac{u_{\Tr{cv}}}{1-\beta^2} ~ &,& ~
  U_{\Tr{cv}}^{\Tr{vc}} + U_{\Tr{vc}}^{\Tr{cv}} = K\epsilon \frac{u_{\Tr{m}}}{1-\beta^2}
\nonumber \\
  U_{\Tr{cc}}^{\Tr{vv}} = K\epsilon \frac{u_{\Tr{e}} e^{i\theta}}{\sqrt{1-\beta^2}} ~ &,& ~
  U_{\Tr{vv}}^{\Tr{cc}} = K\epsilon \frac{u_{\Tr{e}} e^{-i\theta}}{\sqrt{1-\beta^2}}
\nonumber
\end{eqnarray}
where $K = (2\pi)^d / (S_d\Lambda^{\epsilon}m)$. These interactions are diagrammatically represented in the Table \ref{TabVertices}. The RG equations for $\alpha=1$ are:
\begin{eqnarray}\label{RGeq}
\frac{\dd u_\Tr{c}}{\dd l} &=& \epsilon \Bigl\lbrack - u_\Tr{c} - 4u_\Tr{c}^2 - 4u_\Tr{e}^2 \Bigr\rbrack
  \\
\frac{\dd u_\Tr{v}}{\dd l} &=& \epsilon \Bigl\lbrack - u_\Tr{v} - 4u_\Tr{v}^2 - 4u_\Tr{e}^2 \Bigr\rbrack
  \nonumber \\
\frac{\dd u_\Tr{cv}}{\dd l} &=& \epsilon \Bigl\lbrack - u_\Tr{cv} + 2u_\Tr{cv}^2 + 8(1-\beta^2) u_\Tr{e}^2
  \Bigr\rbrack \nonumber \\
\frac{\dd u_\Tr{m}}{\dd l} &=& \epsilon \Bigl\lbrack - u_\Tr{m} - 4u_\Tr{m}^2 + 4u_\Tr{cv}u_\Tr{m} \Bigr\rbrack
  \nonumber \\
\frac{\dd u_\Tr{e}}{\dd l} &=& \epsilon \Bigl\lbrack - u_\Tr{e} + u_\Tr{e} \Bigl(
  -4u_\Tr{c} - 4u_\Tr{v} + 8u_\Tr{cv} - 4u_\Tr{m} \Bigr) \Bigr\rbrack
  \nonumber \\
\frac{\dd e_g}{\dd l} &=& 2e_g - \frac{2u_\Tr{v}}{1-\beta} + \frac{2u_\Tr{cv}}{1-\beta^2} -\frac{u_\Tr{m}}{1-\beta^2}
  \nonumber
\end{eqnarray}

\begin{figure}
\subfigure[{}]{\includegraphics[width=1.6in]{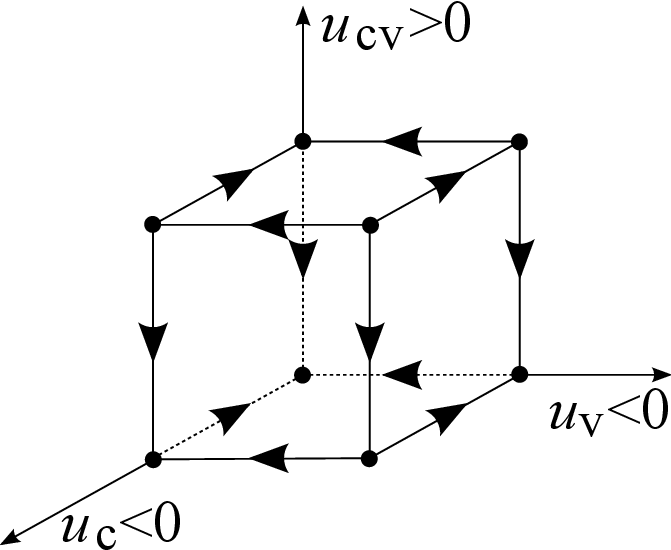}}
\subfigure[{}]{\includegraphics[width=1.6in]{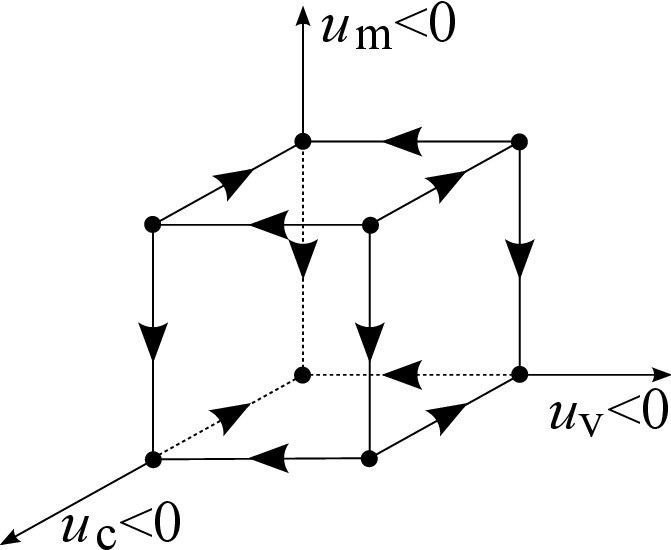}}
\subfigure[{}]{\includegraphics[width=1.6in]{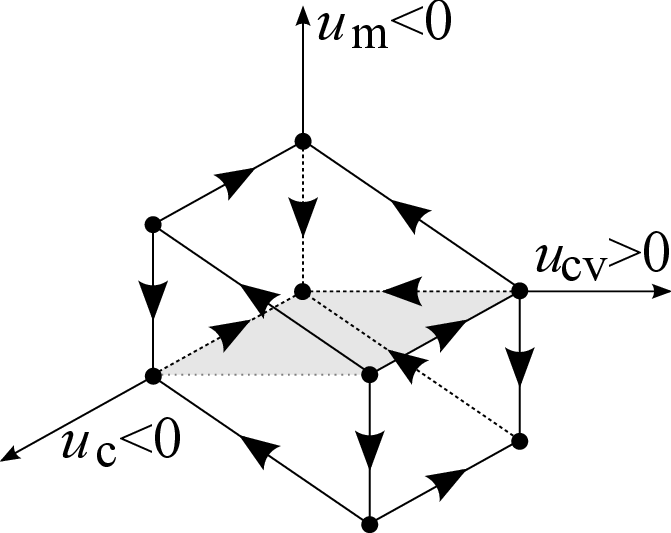}}
\subfigure[{}]{\includegraphics[width=1.6in]{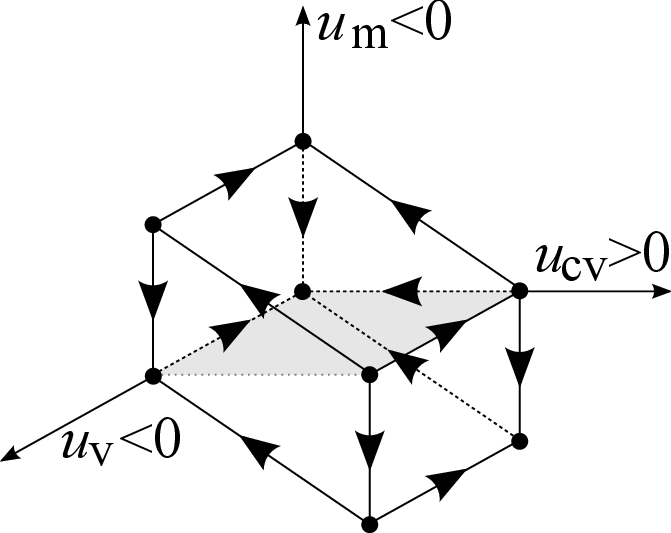}}
\caption{\label{RG2}The fixed points and RG flow of interaction couplings $(u_{\Tr{c}}, u_{\Tr{v}}, u_{\Tr{cv}}, u_{\Tr{m}})$ at $u_\Tr{e}=0$: (a) $u_{\Tr{m}}=0$, (b) $u_{\Tr{cv}}=0$, (c) $u_{\Tr{v}}=0$, (d) $u_{\Tr{c}}=0$. In front of the exposed ``cube'' faces are the run-away regions, signifying pairing correlations in various channels and possible symmetry-broken phases. If one extends the exposed ``cube'' faces in the directions of $u_\Tr{c}>0$, $u_\Tr{v}>0$, $u_\Tr{cv}<0$ and $u_\Tr{m}>0$, the obtained semi-infinite surface (whose corner is a shown ``cube'') encloses the basin of attraction of the Gaussian fixed point.}
\end{figure}

Above two dimensions, there are seventeen fixed points with $\alpha=1$. Sixteen of these fixed points $F_{1}-F_{16}$ are given by all possible combinations of:
\begin{eqnarray}
&& \!\!\!\!\!\! u_\Tr{c}\in\Bigl\lbrace 0, -\frac{1}{4} \Bigr\rbrace ~~ , ~~
  u_\Tr{v}\in\Bigl\lbrace 0, -\frac{1}{4} \Bigr\rbrace ~~ , ~~ u_\Tr{e} = 0 \nonumber \\
&& \!\!\!\!\!\! (u_\Tr{cv},u_\Tr{m}) \in \Bigl\lbrace \Bigl(\frac{1}{2},0\Bigr), \Bigl(\frac{1}{2},\frac{1}{4}\Bigr),
  \Bigl(0,-\frac{1}{4}\Bigr), \Bigl(0,0\Bigr) \Bigr\rbrace \nonumber
\end{eqnarray}
Note that here $u_\Tr{e}$ is always zero. The RG eigenvalues in the subspace of $(u_{\Tr{c}}, u_{\Tr{v}}, u_{\Tr{cv}}, u_{\Tr{m}})$ are $\pm\epsilon$ at all of these fixed points, and allow enumerating $F_{1}-F_{16}$ simply by the variations of relevant/irrelevant flows of $(u_{\Tr{c}}, u_{\Tr{v}}, u_{\Tr{cv}}, u_{\Tr{m}})$. This is illustrated in Fig.\ref{RG2} for the first fifteen fixed points at which at least one of the $u_{\Tr{c}}$, $u_{\Tr{v}}$, $u_{\Tr{cv}}$, $u_{\Tr{m}}$ couplings is zero. Only the Gaussian fixed point is fully stable, while $F_{16}$ with $u_{\Tr{c}}, u_{\Tr{v}}, u_{\Tr{cv}} \neq 0$ and $u_{\Tr{m}}=0$ is fully unstable. The coupling $u_\Tr{e}$ is irrelevant only at the Gaussian fixed point, while it is found to be marginal at $(u_{\Tr{c}}, u_{\Tr{v}}, u_{\Tr{cv}}, u_{\Tr{m}}) \in \lbrace(-1/4,0,0,0),(0,-1/4,0,0),(0,0,0,-1/4)\rbrace$ and relevant otherwise.

\begin{figure}
\includegraphics[width=1.8in]{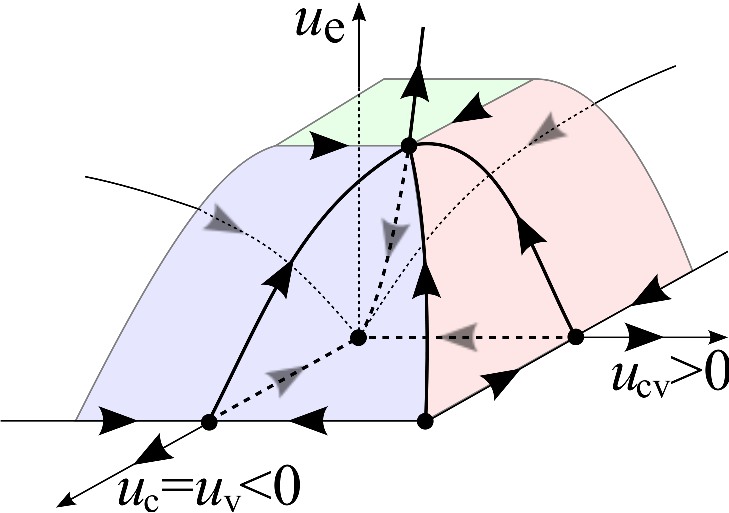}
\caption{\label{RG3}The fixed points and RG flow involving $u_{\Tr{e}} \neq 0$ for $u_{\Tr{m}}=0$, $u_{\Tr{c}}=u_{\Tr{v}}$. The shaded semi-infinite surface encloses the basin of attraction of the Gaussian fixed point.}
\end{figure}

The remaining fixed point $F_{17}$ is the only one with $u_\Tr{e} \neq 0$:
\begin{eqnarray}
&& \!\!\!\!\!\! u_\Tr{e}^2 = \frac{15}{64\left( 11 - 4\beta^2 + 8\sqrt{4\beta^4-7\beta^2+4} \right)} ~~ , ~~ u_\Tr{m}=0
  \nonumber \\
&& \!\!\!\!\!\! u_\Tr{c} = u_\Tr{v} = u_\Tr{cv}-\frac{1}{8} = -\frac{3}{8\left( 5-4\beta^2 + 2\sqrt{4\beta^4-7\beta^2+4}
  \right)} \nonumber
\end{eqnarray}
It has only one relevant direction with RG eigenvalue $\epsilon$, the $u_\Tr{e}$ component being the largest in the corresponding eigenvector. The RG flow in the vicinity of this fixed point is illustrated in Fig.\ref{RG3}. Note that formally there are other solutions stemming from (\ref{RGeq}), but they have $u_\Tr{e}^2<0$ corresponding to time-reversal symmetry violations.

Whenever $u_{\Tr{c}}$, $u_{\Tr{v}}$, $u_{\Tr{cv}}$, $u_{\Tr{m}}$ are relevant, their RG eigenvalue is $\epsilon$, the same as the RG eigenvalue at the unitarity fixed point of a uniform zero-density system which corresponds to vacuum resonant scattering. While this is not surprising when it comes to pairing of two quasiparticles $u_\Tr{c}$ or two quasiholes $u_\Tr{v}$, it is interesting to note that the same resonant scattering interpretation can be applied to the couplings $u_\Tr{cv}$ and $u_\Tr{m}$. We can identify the resonantly scattering quasiparticles by tuning to a fixed point with only one finite coupling and then taking a closer look at the operator corresponding to that coupling. In the case of $u_\Tr{m}$, the operator is
\begin{eqnarray}
f_{\Tr{c}\alpha}^{\dagger} f_{\Tr{v}\beta}^{\dagger}
     f_{\Tr{c}\beta}^{\phantom{\dagger}} f_{\Tr{v}\alpha}^{\phantom{\dagger}} & = &
  -f_{\Tr{c}\alpha}^{\dagger} f_{\Tr{v}\beta}^{\dagger}
     f_{\Tr{v}\alpha}^{\phantom{\dagger}} f_{\Tr{c}\beta}^{\phantom{\dagger}} \nonumber \\
& = & |\Phi_{s}|^2-|\Phi_{t0}|^2-|\Phi_{t\uparrow}|^2-|\Phi_{t\downarrow}|^2 \ , \nonumber
\end{eqnarray}
where the operator $\Phi_s=(f_{\Tr{c}\uparrow}f_{\Tr{v}\downarrow}-f_{\Tr{c}\downarrow}f_{\Tr{v}\uparrow})/\sqrt{2}$ annihilates an interband singlet and the operators  $\Phi_{t\uparrow}=f_{\Tr{c}\uparrow}f_{\Tr{v}\uparrow}$, $\Phi_{t\downarrow}=f_{\Tr{c}\downarrow}f_{\Tr{v}\downarrow}$ and $\Phi_{t0}=(f_{\Tr{c}\uparrow}f_{\Tr{v}\downarrow}+f_{\Tr{c}\downarrow}f_{\Tr{v}\uparrow})/\sqrt{2}$ annihilate triplet pairs. The fixed point(s) at $u_\Tr{m}<0$ can now be associated with the resonant scattering in the interband singlet Cooper channel. Note that the absence of fixed points at $u_\Tr{m}>0$ rules out resonant scattering in the attractive triplet channel (the fixed point at $u_\Tr{cv}=1/2$, $u_\Tr{m}=1/4$ is fully repulsive in the particle-particle channel).

The interaction $u_\Tr{cv}>0$ at its resonant-scattering fixed point is repulsive in the particle-particle channel and cannot lead to a Cooper pair resonance. However, it becomes attractive in the particle-hole channel. Keeping only $u_\Tr{cv}$ finite allows performing a particle-hole transformation in the valence band, after which the theory contains two similarly dispersing fermion fields (particles and holes) in their vacuum states, interacting attractively. Denoting the particle and hole annihilation operators as $f^{\phantom{\dagger}}_\alpha \equiv f^{\phantom{\dagger}}_{\Tr{c}\alpha}$ and $\bar{f}^{\dagger}_{\alpha} \equiv f^{\phantom{\dagger}}_{\Tr{v}\bar{\alpha}}$ respectively, where $\bar{\alpha}$ is the opposite spin of $\alpha$, the $u_\Tr{cv}$ operator can be written as
\begin{eqnarray}
f_{\Tr{c}\alpha}^{\dagger} f_{\Tr{v}\beta}^{\dagger}
     f_{\Tr{v}\beta}^{\phantom{\dagger}} f_{\Tr{c}\alpha}^{\phantom{\dagger}} & = &
  -f_{\alpha}^{\dagger} \bar{f}_{\beta}^{\dagger}
     \bar{f}_{\beta}^{\phantom{\dagger}} f_{\alpha}^{\phantom{\dagger}} \nonumber \\
& = & -|B_{s}|^2-|B_{t0}|^2-|B_{t\uparrow}|^2-|B_{t\downarrow}|^2 \ . \nonumber
\end{eqnarray}
Now the operators $B_s=(\bar{f}_{\uparrow}f_{\downarrow}-\bar{f}_{\downarrow}f_{\uparrow})/\sqrt{2}$, $B_{t\uparrow}=\bar{f}_{\uparrow}f_{\uparrow}$, $B_{t\downarrow}=\bar{f}_{\downarrow}f_{\downarrow}$ and $B_{t0}=(\bar{f}_{\uparrow}f_{\downarrow}+\bar{f}_{\downarrow}f_{\uparrow})/\sqrt{2}$ annihilate singlet and triplet particle-hole pairs. This interaction does not make any distinction between different spins, so that a scattering resonance appears simultaneously in the singlet and all triplet channels. The bound-state resulting from this resonance is an exciton, and symmetry breaking at finite particle and hole density can be either a singlet exciton condensate, or a ferromagnetic state depending on the other couplings as well as higher order terms in the action. In both cases, the present effective theory would favor ordering at zero wavevector, but the circumstances discussed in the following section could lead to antiferromagnetic and other kinds of ordering at finite wavevectors.

The behavior of $u_\Tr{e}$ does not fit this generic resonant scattering picture. Only at the $F_{17}$ fixed point we find the flow of $u_\Tr{e}$ reminiscent of resonant scattering. The absence of other similar fixed points with $u_\Tr{e}\neq 0$, and the fact that the relevant direction at $F_{17}$ is an almost even-amplitude linear combination of multiple couplings, indicate different physics: an ``assisted scattering resonance'' in the Cooper channel between a pair of fermions dynamically resonating between the conduction and valence bands. In fact, assuming $\theta=0$ in (\ref{ScalInt}), a sufficiently strong interaction of this type would give rise to an extended ``sign-changing'' $s$-wave superfluidity in which the pairing gap on the conduction and valence bands has opposite signs. Such an $s^\pm$ pairing is proposed to occur in iron pnictides \cite{Mazin2009}. Other kinds of pairings with different relative phases between the conduction and valence band pairing gaps could be obtained for other values of $\theta$.

The run-away flows in the vicinity of these fixed points are also very important. They indicate the kinds of instabilities of interacting fermions in lattice potentials and circumstances in which they can develop. For electronic systems this information has greater practical use than the detailed properties of the fixed points, because realistic materials can hardly ever be found very close to these fixed points. In generic lattice systems with attractive interactions we find that the favored phases are featureless insulators and superconductors. A singlet superconductor is indicated by the flow of interaction couplings $u_{\Tr{c}}$, $u_{\Tr{v}}$ and $u_{\Tr{m}}$ toward $-\infty$, although as emphasized in the previous section such run-away flows can also produce bosonic Mott insulators in certain cases. Instabilities in the particle-hole channel are discouraged in normal circumstances with attractive microscopic interactions. Even if the interband couplings end up having repulsive character, the generic lattice and microscopic interaction potentials produce relatively small $u_{\Tr{cv}}$ in comparison to $u_{\Tr{c}}$ and $u_{\Tr{v}}$, so that a typical system with attractive interactions in $d+\epsilon$ dimensions flows either to a charge-dynamics influenced insulator state, or toward particle-particle instabilities. With repulsive interactions, however, the same kind of flows near the fixed points featuring $u_{\Tr{cv}}$ take the system either to spin-dynamics influenced insulators, or toward the particle-hole instabilities.

Finding the full structure of fixed points for any $\alpha>|\beta|$ requires allowing all mixing interband couplings to be finite. Preliminary numerical calculations indeed reveal the existence of additional fixed points with finite mixing interactions and $\alpha<1$. However, a systematic search for these fixed points is very difficult due to the large parameter space and the highly non-linear nature of the RG equations that allow $\alpha$ to flow. The details of these fixed points are not crucial for the present discussion and will not be pursued further.

Now we return to the possibility of band inversion which occurs for $\alpha < |\beta|$. First, we note that in normal microscopic circumstances $\alpha$ is close to unity because the intraband couplings $U_n^n$ are larger than the interband ones $U_n^m$, $n \neq m$. The RG flow further accentuates this situation as the flow of all $U_n^m$ is exponential. However, if the interband couplings are large enough in comparison to the intraband ones, we must cure the resulting effective band inversion by identifying the true low energy quasiparticles, which must live at some new crystal wavevectors. The appropriate RG equations need to deal with more than two fermion flavors. Attractive interactions in such strong interband channels would naturally lead to paired states which spontaneously break translational symmetry, while repulsive interactions would give rise to patterned exciton condensates.

\subsection{Transitions involving multiple fermion species, and universality classes}\label{secMF}

The lowest energy quasiparticles in band insulators are often concentrated around multiple symmetry-related wavevectors in the Brillouin zone. For example, the simple cubic periodic potential in three dimensions
\begin{equation}
V(\rv) = 2V \left\lbrack \cos\left( \frac{2\pi x}{a_L} \right) + \cos\left( \frac{2\pi y}{a_L} \right)
  + \cos\left( \frac{2\pi z}{a_L} \right) \right\rbrack \nonumber
\end{equation}
produces a band insulator with two fermions per site (for not too small $V$) whose lowest hole excitations live at $\kv_{\Tr{v}}=(\pi,\pi,\pi)/a_L$ in the valence band and lowest particle excitations live at $\kv_{\Tr{c}1}=(\pi,0,0)/a_L$ and two other symmetry-related wavevectors $\kv_{\Tr{c}2}$, $\kv_{\Tr{c}3}$ in the conduction band \cite{expl1}. An effective fermionic theory of this band insulator requires either one hole or three particle fields for generic pairing transitions of the type discussed in section \ref{secP}. The discussion in section \ref{secPH} has to be extended to one hole and three particle fields in this case.

An effective theory will generally include couplings among all of its fermion fields, and some of the couplings will have the same value by symmetries. As a prototype theory we can take the action (\ref{Seff}) allowing the labels $n,m\dots$ to identify any relevant fermion flavor. Like before, the RG analysis would reveal fixed points and run-away flows corresponding to same-flavor pairing and flavor-mixing instabilities. The latter kind could lead to supersolid phases in the particle-particle channel, or exciton condensates in the particle-hole channel, both bringing translational symmetry breaking and new universality classes. On the other hand, the same-flavor pairing instabilities are the most likely outcome of generic attractive interactions in lattice potentials due to the typically dominant couplings in the same-flavor channel.

All superfluid transitions in the unitarity limit which involve only the same-flavor pairing always belong to the same universality class. This universality class can be characterized by critical ratios between pressure, temperature, energy per particle, and chemical potential (relative to the band edge) at small but finite quasiparticle densities. A useful way of calculating the critical ratios involves applying a Hubbard-Stratonovich transformation to the model (\ref{ContModel1}) to decouple the short-range interaction, and then promoting the obtained two-channel model to an Sp($2N$) symmetry group by introducing $N$ copies of the spinful fermion fields which couple to the same Hubbard-Stratonovich field. Fluctuation corrections to the mean-field thermodynamic functions take the form of $1/N$ expansions, so at least in the limit of large $N$ one can obtain systematic perturbative expressions in the absence of a natural small parameter near unitarity. Taking the physical value $N=1$ and including only the lowest order correction (``Gaussian fluctuations'' of the order parameter) already produces very good estimates in the uniform system \cite{nikolic:033608, Veillette06a}.

Provided that the inter-flavor scattering vanishes at the fixed point, trivial adjustments are needed to accommodate multiple fermion flavors in the presence of a lattice, most notably in the ratios derived from extensive quantities, such as those containing pressure and energy density. For example, the critical pressure $P$ at the finite-temperature $T=T_c$ superfluid transition (in $d=3$)
\begin{equation}
\left. \frac{(P-P_0)/N}{(2m)^{3/2} T^{5/2} n_f} \right|_{T=T_c} = 0.13188 +
\frac{0.4046}{N} + \mathcal{O} (1/N^2) \nonumber
\end{equation}
acquires a factor of $n_f$ in the denominator on the left-hand side, the total number of low-energy particle and hole flavors in the Brillouin zone ($P_0$ is the zero-temperature degeneracy pressure of the band insulator). The value of $n_f$ depends on the bandgap $E_g$; in the limit $T_c \gg E_g$ both particle and hole flavors should be counted in $n_f$, otherwise only particles or holes are important based on the chemical potential.

Another small adjustment of the uniform system $1/N$ expansions in Ref.\cite{nikolic:033608, Veillette06a} is needed in the critical ratios involving the chemical potential. We need to express the chemical potential $\mu$ relative to the nearest band edge. If the conduction band is nearest, then the finite quasiparticle density at zero temperature is obtained when $\mu>0$, while finite density requires $\mu<0$ if the valence band is nearest. The critical temperatures at both $\mu>0$ and $\mu<0$ are universal functions of $|\mu|$:
\begin{equation}
\left. \frac{|\mu|}{T} \right|_{T=T_c} = 1.50448 + \frac{2.785}{N} +
\mathcal{O} (1/N^2) \ . \nonumber
\end{equation}
This expression applies even in the limit $T_c \gg E_g$ when both particles and holes are important, because this limit can be interpreted as $|\mu| \gg E_g$. If the lattice depth is so small that the bandgap closes ($E_g=0$), we must take the larger of the two values of $|\mu|$ obtained by measuring the chemical potential with respect to the overlapping ``conduction'' and ``valence'' band edges. Additional phase transitions below $T_c$ are possible for $T_c \sim |\mu| \gg E_g$, involving the onset of pairing in different channels: particle, hole and interband, each characterized by its own order parameter (see previous section). Re-entrant behavior can be anticipated in this regime when only one fermion species is paired at $T=0$ and another one is separated from the chemical potential by a gap smaller than $T_c$. Then, the thermal population of the fermions across the gap can lead to pairing in additional channels at $0 < T'_c < T < T_c$.

Anisotropy associated with low-energy quasiparticles at symmetry-transforming wavevectors in the Brillouin zone is equally easily treated. For any quasiparticle flavor with dispersion
\begin{equation}
E(\kv) = E_0 + \sum_{i=1}^{d} \frac{k_i^2}{2m_i}
\end{equation}
we redefine momentum so that ${k'}_i^2/2m = k_i^2/2m_i$, where $m$ is a mass to be determined. The measure in path integrals acquires a factor of $\sqrt{(\prod_i m_i)/m^d}$ from this change of variables, which can be absorbed into the redefinition of matter fields. This also leads to a renormalization of all interaction couplings. The choice
\begin{equation}
m = \left(\prod_i m_i\right)^{\frac{1}{d}}
\end{equation}
converts the quasiparticle dispersion into an isotropic one without renormalizing any fields or couplings. It is therefore this geometric mean which should replace the mass in all $1/N$ expansions.

\section{Discussion and conclusions}\label{secDiscussion}

We considered a band insulator subjected to pairing in the unitarity regime as a model system. The simplest realization of such a system is found in trapped neutral ultra-cold gases of alkali atoms placed in an optical lattice. The density of atoms can be chosen to correspond to two atoms per lattice site in the central portion of the trap, while the strength of attractive interactions among them is routinely controlled by the Feshbach resonance. A superfluid transition from a thermally excited band insulator has been already experimentally studied in this kind of a system in the vicinity of the BCS-BEC crossover \cite{Chin2006}.

The focus of our analysis was the characterization of the universal phase diagram featuring $T=0$ transitions between band insulators and superfluid states. In $d>2$ dimensions we identified a BCS limit in which this transition is pair-breaking, meaning that its universal properties are transparently captured by a BCS-like theory. A special limiting case of the pair-breaking transition is found at unitarity, where all interaction effects become independent of microscopic scales, leading to the universal dependence of critical temperature and other thermodynamic functions on the particle density in the superfluid state.

The BEC limit, found at any interaction strength in $d=2$ or at sufficiently strong interactions in $d>2$, brings a different universality class to superfluid transitions. Fermionic excitations belong to high energies, so the effective theory capturing the transition has only bosonic fields. The transition occurs between the superfluid and a bosonic Mott insulator. The universality class is characterized either by the dynamical exponent $z=2$ (generic bosonic mean-field transitions driven by the chemical potential), or $z=1$ (XY transitions driven by interactions at a fixed density).

The BCS and BEC limits considered here are relative to a particular band insulator with a fixed lattice potential and particle density at zero temperature. While the particle density is finite in the ground-state, the unitarity regime between these BCS and BEC limits is found at zero quasiparticle density in the effective low-energy theory describing the band insulator. The full microscopic model includes short-range interactions and multiple fermion bands, with the chemical potential residing in a bandgap. Integrating out high-energy fermions leaves behind the effective theory featuring at most two bands immediately adjacent to the chemical potential (the conduction and valence bands). The remaining low energy fermions experience renormalized interactions, and may exist in multiple flavors as quasiparticles and quasiholes concentrated around different symmetry-related wavevectors in the first Brillouin zone (individually having anisotropic dynamics). All of this complexity reduces to a few relevant interaction couplings in the vicinity of renormalization group (RG) fixed points that signify universal behavior, the most naturally occurring ones corresponding to unitarity in the same universality class as if the system were microscopically uniform.

The physical meaning of these fixed points, revealed by RG, is the resonant scattering of quasiparticles. Multiple flavors of quasiparticles give rise to multiple possibilities for resonant scattering. An interesting discovered possibility is the resonant scattering between particles and holes in the presence of repulsive interactions, the unitarity limit in the particle-hole channel separating the regimes with non-existing and existing exciton bound states (excitonic ``BCS'' and ``BEC'' regimes respectively). Other possibilities not elaborated here also exist in generic circumstances with multiple fermion flavors, leading to translational symmetry breaking in ordered states. Accessing most of these universal regimes requires tuning either the details of lattice potentials, or short-range spatial features (at the lattice spacing scales) of the microscopic interaction potential. This is at least somewhat feasible in trapped gases of cold atoms.

The unitarity regimes in the uniform particle-particle and hole-hole channels can be reached in cold atom systems using Feshbach resonances. The simplest transition is driven by the chemical potential, and hence easy to observe in a trapped gas of cold atoms at an interface between the superfluid and insulating atom clouds. In this case the RG identifies only one relevant interaction parameter, which is tuned by the Feshbach resonance. The transitions driven at fixed density by changing the interaction strength or lattice depth are harder to push to the full unitarity because there are two RG relevant operators (particle-particle and hole-hole scattering lengths) which need to be tuned near their fixed point values. Nevertheless, manifestations of this kind of unitarity can be observed at finite temperatures if the critical temperature is larger than the bandgap.

More challenging is the prospect of observing manifestations of particle-hole resonances. They require strong repulsive interactions between atoms, which can be crafted in a metastable state of a cold atom cloud after a fast magnetic field sweep from the BCS to the BEC limit. The general feasibility of this technique has been demonstrated in an optical lattice \cite{Jordens2008, Schneider2008} and even in continuum \cite{Jo2009}. Similarly challenging is creating resonances between two particles or two holes which live at different wavevectors in the Brillouin zone. Such resonant scattering with a finite momentum transfer cannot be simply induced by external magnetic fields or electromagnetic radiation since neither can deliver or take away large momentum. A trick which might work is to introduce a weak modulation of the optical lattice potential and thus provide a weak umklapp scattering link between the finite momentum open channel, and the usual zero momentum closed channel, the two of which can then be magnetically brought into resonance.

The RG analysis in this paper also provides an indication of the macroscopic properties of states away from the fixed points. If the strength of attractive interactions $U$ is smaller by magnitude than its fixed-point value $|U^*| \propto \epsilon$ in $d+\epsilon$ dimensions, then a gapped fermion system is macroscopically a band insulator. Otherwise, the coupling $U$ flows toward $-\infty$ under RG at \emph{finite length scales}, implying the formation of Cooper pairs at short length scales before the onset of superfluidity at large length scales. It is in this manner that the fermionic RG predicts the existence of bosonic Mott insulators, but a bosonic effective theory is then required to access the superfluid transition at macroscopic scales.

Perhaps the main significance of the presented RG analysis is that the most generic weak-coupling fixed points in fermionic theories, which control the universal properties of insulating and superfluid phases, can be interpreted as resonant scattering. There is a distinction between appropriate ``BCS'' and ``BEC'' regimes in different kinds of pairing channels, in terms of the existence of appropriate two-quasiparticle bound states. The run-away flows of interaction couplings in the ``BEC'' regimes signify the emergence of correlated insulating states separated from ordered phases by transitions in bosonic universality classes. In some circumstances these ``BEC-limit'' insulators may be thermodynamic phases, such as a valence-bond crystal or a spin liquid adjacent to an antiferromagnet (condensate of excitonic ``molecules''), or a charge-density wave adjacent to a superconductor.

Therefore, the presented model and analysis provide a direct insight into the possibilities for the development of strong pairing correlations in fermionic lattice systems. The emergence of boson-dominated superfluid transitions among fermions and the corresponding universality classes can be traced down to the well known physics of BCS-BEC crossovers. Even if interactions are not strong enough to bring the system close to its unitarity limit in empty space, the presence of a lattice frustrates the motion of particles and promotes interaction effects, effectively pushing the system toward its lattice unitarity \cite{Fedichev2004, Koetsier2006}. Furthermore, in two dimensions there is no BCS limit strictly speaking. Two quasiparticles injected into the conduction band will form a bound state regardless of how weak the attractive interactions are. Of course, the size of this ``vacuum'' bound state might be much larger than the spacing between particles, but this does not preclude the bosonic universality of the superfluid transition.

One potentially important aspect of this is that a conceptually similar situation is found in cuprate high temperature superconductors. Cuprates are quasi two-dimensional systems in which the underdoped normal state (pseudogap) exhibits gapped fermionic quasiparticles, albeit with a specific $d$-wave pairing symmetry and a gap of completely different origin than in this paper. A number of unconventional properties of cuprates can be qualitatively understood as being related to a fluctuation-driven transition.

\acknowledgments

I am very grateful to Zlatko Te\v{s}anovi\'{c} for generously sharing his insight, which motivated me to carry out this RG analysis. I also thank Erhai Zhao and Peter Armitage for very helpful discussions. A part of this work was done at the Aspen Center for Physics, and Institute for Quantum Matter at Johns Hopkins University. The support for this research was provided by the Office of Naval Research (grant N00014-09-1-1025A), and the National Institute of Standards and Technology (grant 70NANB7H6138, Am 001).

\appendix

\section{Non-perturbative RG analysis of the run-away flow}\label{app1}

Here we scrutinize in greater detail the run-away RG flow discussed in the section \ref{secP}.

The effective action (\ref{CritTheory1}) contains only gapped particle-like or hole-like excitations, but not both. As a consequence, all Feynman diagrams which contain fermion loops vanish at $T=0$. Each fermion loop represents an internal frequency integral of a product of Green's functions. In the non-relativistic RG, one integrates out the whole Matsubara frequency range $\omega\in(-\infty,\infty)$ at $T=0$, and momenta in a thin shell $|{\bf k}|\in(\Lambda e^{-\Delta l},\Lambda)$, where $\Delta l\ll 1$ and $\Lambda$ is a large momentum cut-off. Since all Green's function poles lie in the same complex half-plane, one can choose to close the complex frequency contour in the other half-plane, where there are no poles, and trivially obtain zero. This is simply a consequence of having a vacuum ground-state. Note that this would not be the case at $T>0$, or in a metallic ground state where the presence of a Fermi surface provides both particle and hole low energy excitations.

\begin{figure}[b]
\includegraphics[width=3in]{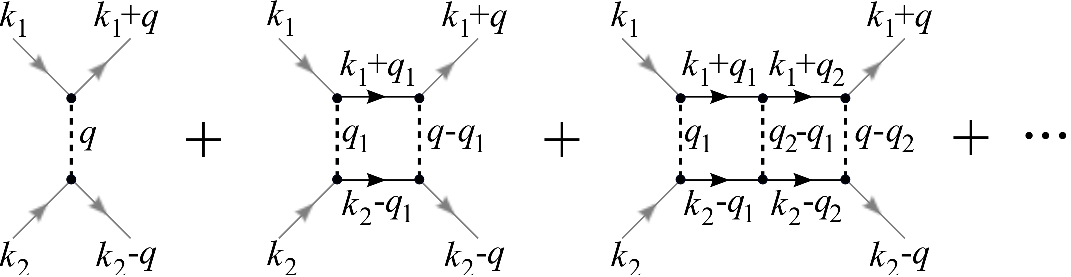}
\caption{\label{Uren}The only non-vanishing diagrams, responsible for renormalizing interactions.}
\end{figure}

As all fermion propagator corrections contain loops, they vanish in our case. The fermion gap $E_0$ flows solely due to rescaling, and $\dd E_0 / \dd l = 2E_0$ in (\ref{RGflow1}) is exact. The only diagrams which do not contain loops, shown in Fig.\ref{Uren}, are responsible for renormalizing interactions. Their geometric sum can be calculated exactly in the weak-coupling limit, when the momentum dependence of interactions and multi-body scattering processes can all be neglected on the basis of being RG-irrelevant. This yields the exact interaction flow in (\ref{RGflow1}) \cite{nikolic:033608, SubirQPT}. We are, however, interested in pushing limits. Below we show that the conclusions of section \ref{secP} hold even in the strong-coupling limit probed by the run-away flow.

The most general short-range two-body interaction consistent with required symmetries is
\begin{equation}\label{Umomentum1}
U(k_1,k_2,k_3,k_4) = U_0^{\phantom{j}} + \sum_{i,j} U_1^{(ij)} \frac{\kv_i \kv_j}{\Lambda^2} +
  \mathcal{O}\left(\frac{k_i^4}{\Lambda^2}\right) \ ,
\end{equation}
where $i,j=1\dots 4$ label the four external lines of the scattering vertex, and $k_i=(\omega_i,\kv_i)$ must add up to zero due to translational invariance. In this notation, the interaction subscript $n$ indicates coupling to $k_i^{2n}$, and $U_0$ is the contact (zero-range) part of interactions used in (\ref{RGflow1}). Similarly, we can consider $n$-body scattering vertices and their momentum dependence. We shall proceed by assuming that all these couplings, except $U_0$, remain small during the accessible part of the run-away flow $U_0 \to -\infty$, and watch for the scale at which this assumption could break down.

We can discover the above scale by dimensional analysis. The full RG equation for $U_0$ contains corrections to (\ref{RGflow1}) of the order of $U_1, U_2\dots$ and various multi-body collision couplings. As long as all such ``irrelevant'' couplings flow toward zero, the relevant run-away flow of $U_0$ is governed precisely by (\ref{RGflow1}). The bare scaling dimensions of $U_n$ are $-(\epsilon+2n)<0$, where $\epsilon=d-2>0$ and $d$ is spatial dimensionality. Similarly, the bare scaling dimensions of multi-body scattering vertices are also finite and negative. These scaling dimensions may be renormalized by dimensionless combinations of $U_0$: the only one available is $\Pi U_0$, where $\Pi$ is the quantity appearing in (\ref{RGflow1}), explicitly calculated below. No ``irrelevant'' coupling can be renormalized by $U_0$ at the tree level, so purely linear terms involving $U_0$ will not appear in the RG equations for $U_1$, $U_2$, etc. There are, of course, quadratic and higher order renormalizations of the ``irrelevant'' terms, but the engineering dimensions require that each appearance of $U_0$ in such corrections be accompanied by a factor of $\Pi$. Therefore, as long as $|\Pi U_0| < 1$, the ``irrelevant'' couplings flow toward zero, dominated by their scaling dimensions, and remain small.

We see that the important scale at which the flow of ``irrelevant'' couplings could be reversed (away from zero) is determined by $|\Pi U_0| \sim 1$. This, however, corresponds to the potential energy of two-body interactions being of the order of $\Lambda^2/2m$, which is the cut-off energy. As discussed in section \ref{secP}, we anyway have to stop the RG program at such scales because the high-energy fermions form Cooper pairs ($U_0<0$) and invalidate the fundamental assumptions behind the present RG. Therefore, the condition $|\Pi U_0| < 1$ holds until the onset of high-energy pairing, and we can safely ignore all irrelevant couplings during the run-away flow of $U_0$.

Below we illustrate this argument by calculating the flow equations which govern the momentum dependence of two-body interactions. Let us calculate the $n^{\textrm{th}}$ diagram $\Pi_{n}(k_{1},k_{2},q)\propto U^{n}$ from Fig.\ref{Uren}:
\begin{widetext}
\begin{eqnarray}\label{Pin1}
&& \!\!\!\!\!\!\!\!\!\!\!\!\!\!\! \Pi_n(k_1,k_2,k_1+q,k_2-q) =
  (-1)^n \int \prod_{i=1}^{n-1} \frac{\dd \Omega_i}{2\pi} \frac{\dd \qv_i}{(2\pi)^d} \;
    \prod_{i=1}^{n-1} \left\lbrack \frac{1}{i(\omega_1+\Omega_i)-E(\kv_1+\qv_i)} \;
    \frac{1}{i(\omega_2-\Omega_i)-E(\kv_2-\qv_i)} \right\rbrack \\
&& \quad \times U(k_1,k_2,k_1+q_1,k_2-q_1) U(k_1+q_1,k_2-q_1,k_1+q_2,k_2-q_2) \cdots
          U(k_1+q_{n-1},k_2-q_{n-1},k_1+q,k_2-q) \nonumber \ .
\end{eqnarray}
\end{widetext}
Since both $\kv_1+\qv_i$ and $\kv_2-\qv_i$ must reside in a thin momentum shell of radius $\Lambda$, the momentum integrals have a significant measure only when $\kv_2=-\kv_1\equiv\kv$. The interaction couplings for other combinations of external momenta receive insignificant renormalization because $\Delta l \ll 1$. Furthermore, the retardation effects (momentum $\omega_i$ dependence of interactions acquired in RG) can be neglected on both physical and formal grounds: the integrated high-energy modes are fast mediators of interactions (with velocities $v\sim\Lambda/m$), and they cannot produce any low-energy poles which would qualitatively affect the frequency integrals. The quantitative effects of retardation on RG flows are of the same kind as those of the ${\kv_i}$-dependence, but less and less dominant as the dimensionality $d$ grows above $2$, which is the dynamical exponent. Having this in mind, we set $\omega_i=0$ and relabel $U(k_1,k_2,k_3,k_4)=U(k,-k,k',-k')\equiv U(\kv,\kv')$. Respecting rotational and time-reversal symmetries, the expression (\ref{Umomentum1}) can be rewritten as
\begin{eqnarray}\label{Umomentum2}
&& \!\!\!\!\!\! U(\kv,\kv') = U_0^{\phantom{j}} + U_1' \frac{\kv^2+\kv'^2}{\Lambda^2}
    + U_1'' \frac{\kv \kv'}{\Lambda^2} + U_2' \frac{\kv^4+\kv'^4}{\Lambda^4} \nonumber \\
&&  + U_2'' \frac{\kv^2 \kv'^2}{\Lambda^4} + U_2''' \frac{(\kv\kv')^2}{\Lambda^4}
    + U_2'''' \frac{(\kv^2+\kv'^2)(\kv\kv')}{\Lambda^4} \nonumber \\
&& + \mathcal{O}\left(\frac{k^6}{\Lambda^6}\right) \ .
\end{eqnarray}
Then, integrating out all Matsubara frequencies $\Omega_i$ in (\ref{Pin1}), and changing variables to $\Qv_i=\kv+\qv_i$ we obtain:
\begin{eqnarray}\label{Pin2}
&& \!\!\!\!\! \Pi_n(\kv,\kv') = (-1)^{n-1} \int \prod_{i=1}^{n-1} \frac{\dd \Qv_i}{(2\pi)^d} \;
     \prod_{i=1}^{n-1} \frac{1}{E(\Qv_i) + E(-\Qv_i)} \nonumber \\
&& \quad \times U(\kv,\Qv_1) U(\Qv_1,\Qv_2) \cdots U(\Qv_{n-1},\kv') \ .
\end{eqnarray}\medskip
where $\Qv_i$ are to be integrated out in the spherical momentum shell so that $E(\pm \Qv_i) = E_0 + \Lambda^2/2m$. Note that $\Pi_1(\kv,\kv') = U(\kv,\kv')$. Now we substitute (\ref{Umomentum2}) here and collect the terms at various powers of momenta. This can be done systematically in a fashion which is perturbative only in the assumed small couplings $U_1,U_2\dots$ but not in $U_0$. Namely, the last expression factorizes if we neglect the momentum dependence $U(\kv,\kv')\to U_0$, making the sum over $\Pi_n$ a simple geometric progression. Since the renormalized interactions remain short-ranged, we can also expand (\ref{Pin2}) in powers of momenta for any $n\ge 2$:
\begin{widetext}
\begin{eqnarray}\label{Pin3}
\Pi_n(\kv,\kv') &=& (-\Pi \Delta l)^{n-1} \Biggl\lbrace U_0^n
    + U_1' U_0^{n-1} \left\lbrack 2(n-1) + \frac{\kv^2+\kv'^2}{\Lambda^2}\right\rbrack
    + U_2' U_0^{n-1} 2(n-1) \\
&&  + U_2'' U_0^{n-1} \left\lbrack n-2 + \frac{\kv^2+\kv'^2}{\Lambda^2}\right\rbrack
    + U_2''' U_0^{n-1} \left\lbrack \frac{n-2}{d} + \frac{\kv^2+\kv'^2}{d\Lambda^2}\right\rbrack \Biggr\rbrace
    + \mathcal{O}\left( U_1^2, U_2^2, U_3, \frac{\kv^4}{\Lambda^4} \right) \ . \nonumber
\end{eqnarray}
\end{widetext}
where
\begin{eqnarray}
&& \Pi = \frac{1}{\Delta l} \int \frac{\dd^d \Qv}{(2\pi)^d} \; \frac{1}{2E(\Qv)} =
         \frac{S_d}{(2\pi)^d} \; \frac{m\Lambda^d}{2mE_0 + \Lambda^2} \nonumber \\
&& \;\;\; \xrightarrow{\Lambda^2\gg 2mE_0} \frac{S_d}{(2\pi)^d} m \Lambda^{d-2}
\end{eqnarray}
is the positive constant in (\ref{RGflow1}), and $S_d$ is the $d$-dimensional unit-sphere area.

Note that the fermion gap $E_0$ also grows under RG, and can in principle inflate to the cut-off scale and reduce the value of $\Pi$. The most adequate way to handle this would be to fix a cut-off energy $E_\Lambda$ and derive the momentum cut-off $\Lambda$ from $E_\Lambda = E_0 + \Lambda^2/2m$. Then, $\Lambda$ decreases due to the scaling of $E_0$, pulling $\Pi$ toward zero as $E_0 \to E_\Lambda$. Should that happen before $U_0$ escalates, we would have to abort the RG program with a conclusion that the system is a band-insulator: further RG flow is halted since no more particle modes exist below the cut-off. However, if the system is instead a bosonic Mott insulator of Cooper pairs, or a superconductor, the interaction $U_0$ will be the first to reach the cut-off scale while $E_0$ is still far enough from the cut-off to keep $\Pi$ roughly a constant for qualitative purposes.

We can now sum up exactly all $\Pi_n$, evaluated with $U(l)$ at a scale parametrized by $l$. This gives us the coarse-graining contribution to the renormalization of interaction couplings:
\begin{widetext}
\begin{eqnarray}\label{UU1}
U(l+\Delta l) &=& \sum_{n=1}^{\infty} \Pi_n(\kv,\kv') = \frac{U_0}{1 + \Pi U_0 \Delta l} \left\lbrack
  1 - \frac{\left( 2 U_1' + 2 U_2' + U_2'' + U_2'''/d \right) \Pi \Delta l}{1 + \Pi U_0 \Delta l}
    - U_2'' - \frac{U_2'''}{d} \right\rbrack + U_2'' + \frac{U_2'''}{d} \nonumber \\
&&  + \frac{\kv^2+\kv'^2}{\Lambda^2} \; \left\lbrack \frac{U_1' + U_2'' + U_2'''/d}{1 + \Pi U_0 \Delta l}
    - \left( U_2'' + \frac{U_2'''}{d} \right) \right\rbrack
    + U_1'' \frac{\kv\kv'}{\Lambda^2} + \mathcal{O}\left( U_1^2, U_2^2, U_3, \frac{\kv^4}{\Lambda^4} \right) \ .
\end{eqnarray}
\end{widetext}
After substituting (\ref{Umomentum2}) on the left-hand side and multiplying both sides by $1 + \Pi U_0 \Delta l$, we are ready to take the $\Delta l \to 0$ limit and separate the terms proportional to different powers of momenta. Then, we add the scaling contributions to RG flows, which depend on the power of momentum (the bare scaling dimension of $U_n$ is $-\epsilon-2n$). Neglecting $U_1^2, U_2^2, U_3$ and higher order renormalizations, the final RG equations for interactions up to $\mathcal{O}(k^2/\Lambda^2)$ are:
\begin{eqnarray}
&& \!\!\!\!\! \frac{\dd U_0}{\dd l} = -\epsilon U_0 - \Pi U_0^2
      - \Pi U_0 \left(2U_1' + 2U_2'\right) + \cdots \nonumber \\
&& \!\!\!\!\! \frac{\dd U_1'}{\dd l} = -\left(2+\epsilon+\Pi U_0\right) U_1'
      - \Pi U_0 \left(U_2'' + \frac{U_2'''}{d}\right) + \cdots
   \nonumber \\
&& \!\!\!\!\! \frac{\dd U_1''}{\dd l} = -\left(2+\epsilon\right) U_2'' + \cdots \\
&& \cdots \cdots \cdots \nonumber
\end{eqnarray}
These flow equations are correct to all powers of $U_0$ and formally take into account the full $\Pi(E_0)$ dependence. Their properties have been discussed earlier in this appendix. The couplings $U_1, U_2\dots$ are effectively irrelevant even during a run-away flow of $U_0$, until the high-energy fermions form Cooper pairs, so that we may think of the equations (\ref{RGflow1}) as being ``exact''.

%\bibliographystyle{/usr/share/texmf/bibtex/bst/revtex/prsty}
%\bibliography{/home/dasko/Science/Bibliography/references}

\end{document}